# Electron-phonon coupling in metals at high electronic temperatures

**N. Medvedev[1,2,*], I. Milov[3]**

1. Institute of Physics, Czech Academy of Sciences, Na Slovance 2, 182 21 Prague 8, Czech Republic
2. Institute of Plasma Physics, Czech Academy of Sciences, Za Slovankou 3, 182 00 Prague 8, Czech Republic
3. Industrial Focus Group XUV Optics, MESA+ Institute for Nanotechnology, University of Twente, Drienerlolaan 5, 7522 NB Enschede, The Netherlands

## Abstract

Electron-phonon coupling, being one of the most important parameters governing the material evolution after ultrafast energy deposition, yet remains the most unexplored one. In this work, we applied the dynamical coupling approach to calculate the nonadiabatic electron-ion energy exchange in nonequilibrium solids with the electronic temperature high above the atomic one. It was implemented into the tight-binding molecular dynamics code, and used to study electron-phonon coupling in various elemental metals. The developed approach is a universal scheme applicable to electronic temperatures up to a few electron-Volts, and to arbitrary atomic configuration and dynamics. We demonstrate that the calculated electron-ion (electron-phonon) coupling parameter agrees well with the available experimental data in high-electronic-temperature regime, validating the model. The following materials are studied here – fcc metals: Al, Ca, Ni, Cu, Sr, Y, Zr, Rh, Pd, Ag, Ir, Pt, Au, Pb; hcp metals: Mg, Sc, Ti, Co, Zn, Tc, Ru, Cd, Hf, Re, Os; bcc metals: V, Cr, Fe, Nb, Mo, Ba, Ta, W; diamond cubic lattice metals: Sn; specific cases of Ga, In, Mn, Te and Se; and additionally semimetal graphite and semiconductors Si and Ge. For many materials, we provide the first and so far the only estimation of the electron-phonon coupling at elevated electron temperatures, which can be used in various models simulating ultrafast energy deposition in matter. We also discuss the dependence of the coupling parameter on the atomic mass, temperature and density.

## I. Introduction

The field of high-energy-density in matter has been developing for a few decades, especially since the advent of the femtosecond lasers [1]. High-power laser technology is one of the most widely used nowadays, with applications in fundamental research [2], materials processing and micromachining [3,4], biology and medicine [5,6], and many others [7]. Despite this rich history of lasers use and research, many fundamental aspects of material response to ultrafast irradiation remain poorly understood [8].

After ultrafast high-energy-density deposition, matter enters a nonequilibrium state, which then proceeds to relax towards thermal equilibrium. Excited electronic ensemble equilibrates at an electronic temperature high above the atomic one. It induces a variety of effects, including electronic energy exchange with atoms/ions, modification of the interatomic potential, possible charge nonequilbirium and Coulombic response of the ions, evolution of the material properties and phase transitions [9–11]. Within this work, we will only focus on the first one, the electron-ion coupling.

Since the 1950s, it became a common practice to describe evolution of matter after high-energy-density deposition within the so-called two-temperature model (TTM). The model was introduced by Shabansky and Ginzburg [12], and independently by Lifshitz, Kaganov and Tanatarov [13]. Its widespread use in the laser-matter interaction community started after the works by Anisimov *et al*. [14], and the model remains one of the most commonly used computational tools in the community to this day [9]. The popularity of the TTM due to its simplicity spiked again after its combination with molecular dynamics simulations demonstrated the power of such a combined approach [15].

[*] Corresponding author's email: nikita.medvedev@fzu.cz

Despite the widespread use of the TTM, one of the key parameters – the electron-phonon coupling strength – remains elusive and is the least known parameter in the model [9,16]. The calculations at elevated electronic temperatures are only available for a few metals, and are in a high demand [16–18]. However, recent experiments started to indicate that the model calculations based on the standard extension of the Eliashberg spectral function formalism to a finite electronic temperature might overestimate the coupling strength [19,20].

Here we present a nonperturbative extension of a recently proposed approach [21] to calculation of the electron energy coupling to the ionic motion. The previously developed formalism that we refer to as the 'dynamical coupling' is extended here to nonorthogonal Hamiltonians. We then apply it to various elemental metals across the periodic table, semimetal graphite and semiconductors Si and Ge. We start by demonstrating its good agreement with the available experimental data on electron-phonon coupling at high electronic temperatures. We proceed by presenting our predictions for the coupling strength for materials, for which no experimental data in a two-temperature regime and often no theoretical estimations are known. We also identify a few cases where our results disagree with the experimental ones, indicating potentially interesting directions of future research. At last, we discuss the dependence of the coupling strength on other parameters, such as the atomic temperature and the material density, and identify some trends of the coupling parameter for materials across the periodic table.

In the context of our calculations, it is more appropriate to use a more general term electron-*ion* coupling, as opposed to a more common term electron-*phonon* coupling, since we are not restricted to harmonic motion for ions (phonon modes), as will be described below. For simplicity, we use the latter term in most of the cases, unless it is important to emphasize the difference.

## II. Model

XTANT-3 code is a hybrid approach developed for modeling of a material response to femtosecond irradiation [11]. It combines on-the-fly the following modules within a unified model: (a) a Monte Carlo (MC) model for photoabsorption and nonequilibrium kinetics of high-energy electrons; (b) rate equations for low-energy electron evolution on the transient band structure; (c) Boltzmann collision integrals coupling low-energy electrons to the atomic motion; (d) a transferable tight binding (TB) formalism for a description of the transient band structure and the atomic potential energy surface; (e) molecular dynamics (MD) tracing of atomic motion.

Within this work, we do not trace the nonequilibrium electronic stage, and thus the MC model serves only as a means of delivering energy into the electronic system, elevating the electronic temperature. To fulfill this condition, we use photons with a low energy (10 eV), so there are no high-energy electrons produced. The low-energy fraction of electrons is assumed to follow the Fermi-Dirac distribution, $f(E) = 2(1 + \exp((E - \mu)/T_e))^{-1}$, at all times (here $\mu$ is the chemical potential of electrons, $T_e$ is the electronic temperature, and the factor of 2 accounts for the spin degeneracy).

The Boltzmann collision integral assumes the following form for a coupling of degenerate electrons to atoms [21]:

$$I_{e-a}^{ij} = w_{ij} \begin{cases} f(E_i)\left(2 - f(E_j)\right) - f(E_j)\left(2 - f(E_i)\right)e^{-E_{ij}/T_a}, \text{for i > j} \\ f(E_j)\left(2 - f(E_i)\right)e^{-E_{ij}/T_a} - f(E_i)\left(2 - f(E_j)\right), \text{otherwise} \end{cases} \quad (1)$$

here $w_{ij}$ is the rate of electron transitions triggered by atomic motion, $E_{ij} = E_i - E_j$ is the difference between the energies of the two levels, and $T_a$ is the atomic temperature. The exponential factor arises due to the integrated Maxwellian distribution of atoms, which is obtained from the MD simulation at each timestep (note that the original expression in [21] was incorrect). This shape of the Eq.(1) ensures that the detailed balance is fulfilled, which is not always the case in nonadiabatic molecular dynamics simulations (see e.g. a discussion in [22]).

The dynamical coupling approach to calculation of the transition rates [21] is extended here to include nonperturbative matrix elements and generalized to non-orthogonal basis sets. The full transition probability between the eigenstates $i$ and $j$ is $W_{ij} = \left|\langle\psi_j(t)|\psi_i(t_0)\rangle\right|^2$, considering that the Hamiltonian of the system is constant within each timestep of molecular dynamics and only suddenly changes between the consecutive steps [21]. In contrast to our previous works [21,23], here we do not use the first order approximation, because, in fact, the tight binding molecular dynamics (TBMD) provides the eigenfunctions of the Hamiltonian on each timestep. Thus, the transition rate can be generally calculated as follows:

$$\begin{aligned} w_{ij} = \frac{dW_{ij}}{dt} &\approx 2\left|\langle\psi_j(t)|\psi_i(t_0)\rangle\right| \frac{\left|\langle\psi_j(t)|\psi_i(t_0)\rangle - \langle\psi_j(t_0)|\psi_i(t_0)\rangle\right|}{\delta t} \\ &= 2\frac{\left|\langle\psi_j(t)|\psi_i(t_0)\rangle\right|^2}{\delta t} \end{aligned} \quad (2)$$

where the derivative is approximated with the finite difference method for the molecular dynamics timestep $\delta t$, and the wave-functions, $\psi(t)$, are calculated correspondingly on two consecutive steps: $t_0$ and $t=t_0+\delta t$. Here we took into account the orthogonality of the eigenstates within the same timestep.

Using the LCAO basis set within the tight binding Hamiltonian, $\psi_i = \sum_\alpha c_{i,\alpha}\varphi_\alpha$, Eq.(2) can be written in the following manner:

$$w_{ij} = 2\sum_{\alpha,\beta} \frac{\left|\langle c_{i,\alpha}\varphi_\alpha(t)|c_{j,\beta}\varphi_\beta(t_0)\rangle\right|^2}{\delta t} \approx \frac{2}{\delta t}\sum_{\alpha,\beta}\left|c_{i,\alpha}(t)c_{j,\beta}(t_0)S_{\alpha,\beta}\right|^2 \quad (3)$$

Here $S_{\alpha,\beta}$ is the overlap matrix. Strictly speaking, the overlap matrix must be calculated as the overlap between the basis functions on the two consecutive timesteps, however we approximate it as the TB overlap matrix on the current timestep $t$. For sufficiently small MD steps, this is a satisfactory approximation. Note that in a case of an orthogonal Hamiltonian ($S_{\alpha,\beta} = \delta_{\alpha,\beta}$), using the split-step method of the finite difference derivative calculations reduces it to the previously used scheme [21], except for the factor of 2.

This expression, combined with the Boltzmann collision integral (1) implemented into the TBMD, has a linear dependence on the numerical timestep. To eliminate this dependence, we introduced a factor of $2e/(\hbar\delta t)$ ($e$ is the electron charge and $\hbar$ is the Planck's constant providing the dimensionality of time consistent with that of the MD timestep to render the multiplier dimensionless). Thus, we implemented the following expression into the XTANT-3 code:

$$w_{ij} = \frac{4e}{\hbar\delta t^2}\sum_{\alpha,\beta}\left|c_{i,\alpha}(t)c_{j,\beta}(t_0)S_{i,j}\right|^2 \quad (4)$$

The multiplier is rather *ad hoc* and is introduced here for numerical reasons, however, we will demonstrate below that it provides a reasonable agreement with experiments. We emphasize here that it is not an adjustable parameter in the model, and Eq.(4) does not change in all further simulations. The reported scheme for the electron-ion coupling thus serves as a simple and straightforward method of accounting for nonadiabatic effects in tight binding MD. Note that the developed scheme does not assume harmonic motion for atoms (phonon modes), nor does it rely on the atomic system to be in a crystalline phase (periodic structure). It is, in principle, applicable to any kind of atomic motion and material structure, including e.g. melted and amorphous matter.

Knowing the electron-ion collision integral, Eq.(1), we can define the coupling parameter [24]:

$$G(T_e) = (T_e - T_a)^{-1} \sum_{i,j} E_j I_{e-a}^{ij} \tag{5}$$

In this work, we use the NRL transferable tight binding parameterization [25]†. It is one of the best and thoroughly tested transferable tight binding parameterizations, reliable for description of the electronic band structure and atomic properties of metals and some semiconductors. It is a nonorthogonal parameterization based on $sp^3d^5$ LCAO basis set. In certain cases, we had to add a pair-wise short range repulsive potential to make the molecular dynamics stable at elevated temperatures and pressures [26].

For each material, we run ten XTANT-3 simulations with different initial conditions and parameters of irradiation (various pulse durations and deposited doses). Prior to irradiation, the system is thermalized during 300-500 fs, starting from the ideal crystal atomic positions and assigning random velocities to atoms in accordance with the Maxwell distribution at the room temperature. The irradiation is made with 10 eV photons, for which case there is no nonequilibrium electron cascade effects [27], thus the irradiation only increases the electronic temperature within a certain timeframe. The pulse duration is varied between 10 and 60 fs, and the deposited dose between 3 and 4 eV/atom. Such a dose elevates the electronic temperature to some 20-30 kK, depending on a material. The calculations are carried out until the irradiation is over, i.e. for approximately two pulse durations. Variations of the pulse duration, the irradiation instant (pulse intensity peak position) and the dose are made to exclude possible artificial correlations effects. Such variations guarantee that different electronic temperatures are reached at different times, and hence at different positions and velocities of atoms. These improves reliability of our final averaged results.

Due to extremely short irradiation durations and hence a short calculation time, an increase of the atomic temperature is minor throughout the simulations. We use an NVE (microcanonical) ensemble, periodic boundary conditions, and a timestep of 1 fs in our simulations. Convergence of the results with respect to the numerical parameters is analyzed in the Appendix.

Knowing the time evolution of the system under electronic excitation, we then extract the coupling parameter as a function of the electronic temperature, and average over the ten simulation runs. In all the plots below, the average coupling parameter is shown with the error bars corresponding to the mean standard deviation in the ten simulation runs.

## III. Results

We start by comparing our model with recently available experimental data on the electron-phonon coupling in materials with a highly excited electronic system. After that, we proceed by presenting the calculated data for materials for which only low-temperature or no data are available. Those materials are presented in groups according to their space symmetries. Within each group, the materials are sorted by the increasing atomic number.

† The NRL TB parameterizations are available online at: http://esd.cos.gmu.edu/tb/tbp.html

Wherever possible, we compare our calculations with other models, such as calculations of the electron-phonon mass enhancement parameter, λ [28–30], and its electron-temperature-dependent extensions [16]. We convert the electron mass enhancement parameter into the electron-phonon coupling parameter with the following relation [16,29]:

$$G_0 = G(T_e = 0\ K) = \pi \hbar k_B \lambda \langle \omega^2 \rangle N(\varepsilon_F) \quad (6)$$

here $k_B$ is the Boltzmann constant, $\langle \omega^2 \rangle$ is the second moment of the phonon spectrum, and $N(\varepsilon_F)$ is the electronic density of states at the Fermi level [16,29]. To define the second moment of the phonon spectrum, we use the approximation from Ref. [31]:

$$\hbar^2 \langle \omega^2 \rangle \approx (0.68 k_B \theta_D)^2 \quad (7)$$

with $\theta_D$ being the Debye temperature of the material.

We emphasize that the quantity $N(\varepsilon_F)$ in some materials where the density of states (DOS) has sharp peaks at the Fermi level is very sensitive to the calculation details and parameters, and thus estimations of the coupling parameters with Eq.(6) may be rather uncertain [31]. That is why, in what follows, the plotted data from different authors may exhibit large discrepancies. Despite our best efforts to collect as much experimental and theoretical data for comparisons as possible, the reference list is probably not exhaustive.

We note that the calculations within XTANT-3 model are performed for the single gamma-point in the Brillouin zone, and thus may not describe a low electronic temperature regime accurately. The data shown for electronic temperatures below some ~1000-2000 K may be expected to deviate from the experimental ones and other models in some cases.

### III.1. Model validation

We compare the calculated electron-ion coupling parameter with available experimental data at high electronic temperatures for Al and Au. In our XTANT-3 simulations, a supercell is composed of 4x4x4 conventional unit cells (256 atoms). A comparison with experiments and other simulations for aluminum is shown in Figure 1. We can see that the XTANT-3 calculations agree reasonably well (within the experimental error bars) with the experimental data of Waldecker *et al.* [19] at finite electronic temperatures available up to 4000 K. In other experiments, the electron-phonon coupling was measured only at the room temperature, and the results significantly vary.

Several calculations were previously performed utilizing various theoretical approaches. As one can see, the results differ significantly, although with a common trend of increasing electron-phonon coupling with increasing electron temperature. The curves marked as "Gorbunov 1" and "2" [24] are obtained via dynamic structure factor formalism, which assumes electrons wave functions to be plane waves of free particles within the first Born approximation. The curve "Gorbunov 2", being closer to the room temperature experiment values [19,28,29,32–34], additionally assumed increased screening. Waldecker *et al.* [19] and Brown *et al.* [18] used finite-electron temperature extensions of the electron mass enhancement parameter, similar to the work of Lin *et al.* [16], but with the phonon spectrum calculated with the DFT method. Without additional adjusting of the mass enhancing parameter, their results overestimate the experimental data. Lin *et al.* reported that they used the experimental data point by Hostetler *et al.* [33] to fit their mass enhancement parameter, thereby reproducing the experimental data. Muller *et al.* [35] and "Gorbunov 3" [24] used the full electron-phonon Boltzmann collision integral with a simplified expression for the matrix element (within the jelly model) and the free-electron approximation. Petrov *et al.* [17] used a similar approach, but with effective electron masses different for *s* and *d* electrons, and with Lindhard screening. As one can see, several calculations provide a reasonable agreement with the experimental data of Waldecker *et al.* [19].

Note that some authors adjusted their calculations to the room-temperature experimental data for aluminum (e.g. [16,24]), while unadjusted *ab-initio* calculations within the Eliashberg formalism produce overestimated results [18,19]. It has been suggested that a solution to this discrepancy is the fact that electrons should couple differently to different phonon modes (longitudinal and transversal), accounting for which reduced the coupling strength [19] (not shown in Figure 1). The dynamical coupling scheme reported here naturally accounts for this effect within the XTANT-3 and produces the coupling strength close to the experimental one.

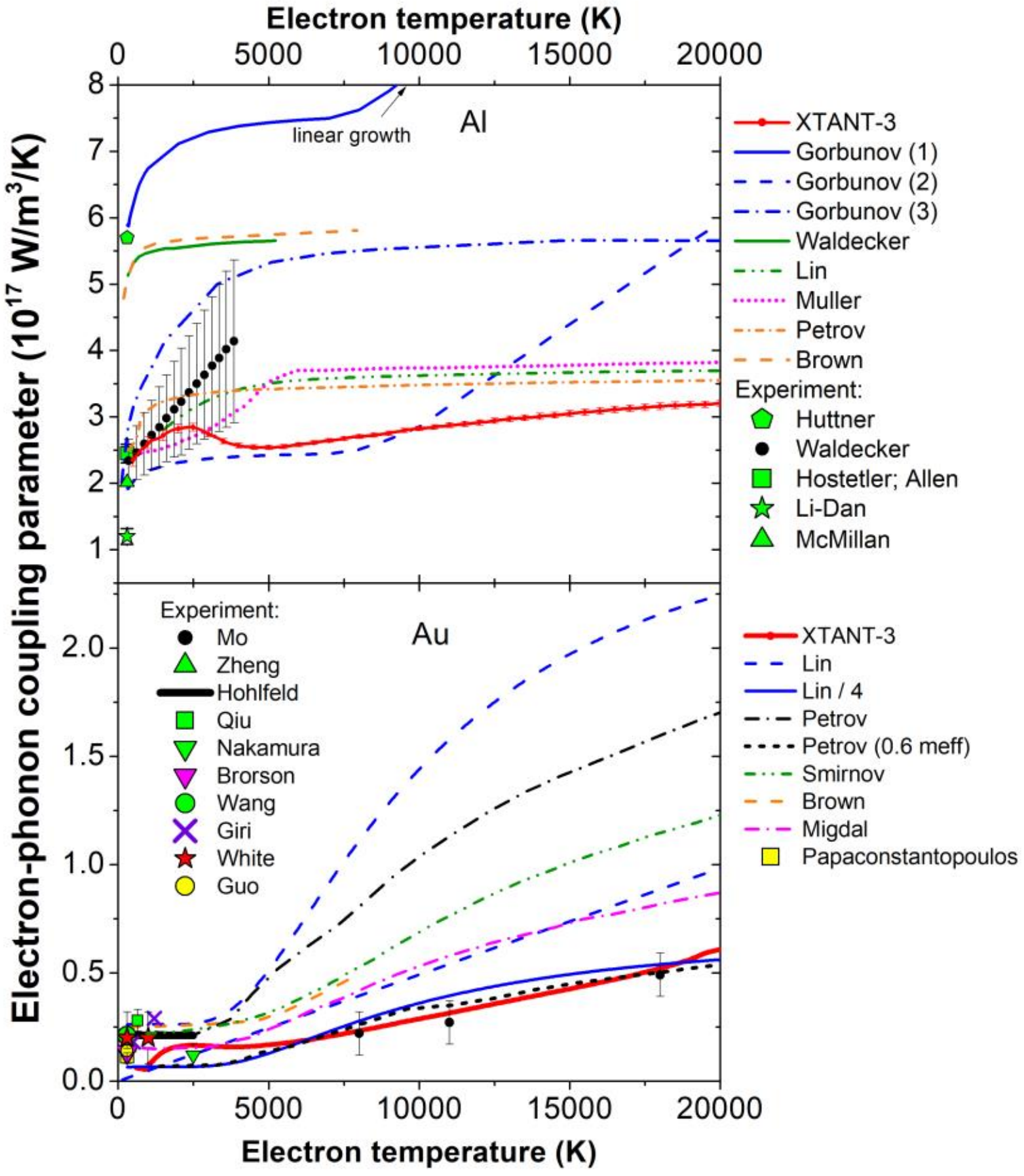

*Figure 1. Top panel: Electron-phonon coupling parameter in* ***aluminum*** *calculated within XTANT-3 dynamical coupling approach, compared with various theoretical and experimental data available. Experimental data include those of Huttner et al.* [32]*, Waldecker et al.* [19]*, Hostetler et al.* [33]*, and Li-Dan et al.* [34]*. Theoretical estimates are by Gorbunov et al.* [24] *(three different models were reported in there), Lin et al.* [16]*, McMillan* [29]*, Allen et al.* [28]*, Brown et al.* [18]*, Petrov et al.* [17]*, Muller et al.* [35]*, and Waldecker et al.* [19]*. Bottom panel: Electron-phonon coupling parameter in* ***gold*** *calculated within XTANT-3 dynamical coupling approach, compared with various theoretical and experimental data available. Experimental data include those of Mo et al.* [20]*, Hohlfeld et al.* [36]*, Qui et al., Nakamura et al.* [37]*, Brorson et al.* [38]*, Zheng et al.,* [39]*, Giri et al.* [40]*, White et al.* [41]*, and Guo et al.* [42]*. Theoretical estimates are by Lin et al.* [16] *(and the same data rescaled by the factor of 1/4), Brown et al.* [18]*, Smirnov et al.* [43]*, Migdal et al.* [44]*, Wang et al.* [45]*, Papaconstantopoulos* [30]*, and two curves from Petrov et al.* [17]*.*

A comparison of XTANT-3 results with experimental data and other calculations for gold is shown in the bottom panel of Figure 1. There, again, we see that our results agree with the experimental points at high electronic temperatures [20], whereas other works overestimate it. We note that calculations by Lin *et al*. [16], which are now widely used in the community, overestimate the experimental data by a factor of four: indeed, rescaled by 1/4, the calculated curve goes through the experimental data. Calculations by Petrov *et al.* [17] also overestimate the experimental data, unless a reduced effective electron mass is used ($m_{eff}$ = 0.6 $m_e$ reported in the original work). Brown *et al*. calculated a slower increasing coupling, as they included dependencies of several parameters entering the coupling strength on the electron temperature [18]. The same methodology was used by Smirnov *et al.* [43]. XTANT-3, including evolution of the atomic potential (and, correspondingly, the phonon spectrum) and the electron hopping matrix elements, predicts even slower increase of the coupling strength.

The experimental data by Mo *et al*. [20] are plotted against the maximal electronic temperature reached in the experiment. The authors estimate that the coupling parameter stays nearly constant during the entire experiment, defined by the value reached at the maximal electronic temperature. They note that the experimental data on the transient evolution of the atomic displacements are best reproduced within the TTM by assuming a constant coupling parameter [20], which agrees with our calculations that show only a slow increase of the coupling parameter with the increasing electronic temperature.

The comparison with the experimental data currently available in the literature validates our approach at high electronic temperatures, at least for the two elemental fcc metals. To the best of our knowledge, there is no data on the electron-phonon coupling strength published for other materials in the high-electron-temperature regime as of yet.

### III.1. Fcc metals

In this section, we present the results for other fcc metals, namely: Cu, Ca, Ni, Sr, Y, Zr, Rh, Pd, Ag, Ir, Pt, Pb. We use the supercell composed of 4x4x4 (Cu, Rh, Pd, Ag, Ir, Pt) or 5x5x5 (Ca, Ni, Sr, Y, Zr, Pb) conventional orthogonal unit cells (256 or 500 atoms, correspondingly).

Figure 2 shows the calculated coupling parameter in copper, calcium, nickel and strontium. In copper, XTANT-3 data agree with the most low-electron-temperature experimental data, but again show a slower increase with $T_e$ than other models, such as Lin's *et al*. [16] and Brown's *et al.* [18]. However, more recent and advanced Brown's et al. calculations demonstrate a slower increase than Lin's et al., closer to that of XTANT-3. The recent results by Migdal et al. from [53] differ from the older results from [44] by the fact that newer ones included the changes of the density of states with increase if the electronic temperature. We also note that a simple rescaling of Lin's results does not reproduce our results, implying that there might be influence of some effects on the coupling that were not accounted for in earlier works, such as an effect of the modification of the interatomic potential under electronic excitation (known as phonon hardening in bulk metals [46]). As we will also see later, such an effect leads to an increase of the electron-phonon coupling, which seems to be the case here at high electronic temperatures ($T_e$ > 10000 K), where Eliashberg-formalism-based results show a saturation of the coupling with increasing $T_e$, whereas XTANT-3 predicts an increase.

Unfortunately, as of now, there is no experimental data to validate the model at high electronic temperatures in copper. Cho's *et al*. results [52] on the coupling in the warm dense matter state of copper do not correspond to the conditions simulated in XTANT-3, where the atomic system was in fcc structure at room temperature. Cho *et al*. reported that their system is best described as a liquid copper at elevated temperatures [52], due to the long timescales of their experiment. In fact, different points shown correspond to different atomic temperatures reached at different times after irradiation with different fluences. Thus, they are shown here merely as a limiting case, and should not be directly compared to the XTANT-3 results. For an illustration, we calculated a coupling parameter in copper for the atomic temperature of 2000 K, which is above the melting temperature of copper, $T_{melt}$ = 1358 K at normal conditions. The calculated coupling parameter goes through one set of three experimental points that are measured at similar conditions. The other sets of experimental points are measured at different fluences, thus corresponding to different atomic temperatures (and perhaps densities) of the target. We see from Figure 2a that a careful analysis of the experimental data is required to interpret the measurements, since the coupling parameter is not a function of a single variable – the electronic temperature – but depends on many other parameters, such as the atomic temperature and density. More on this point will be shown in the Discussion section.

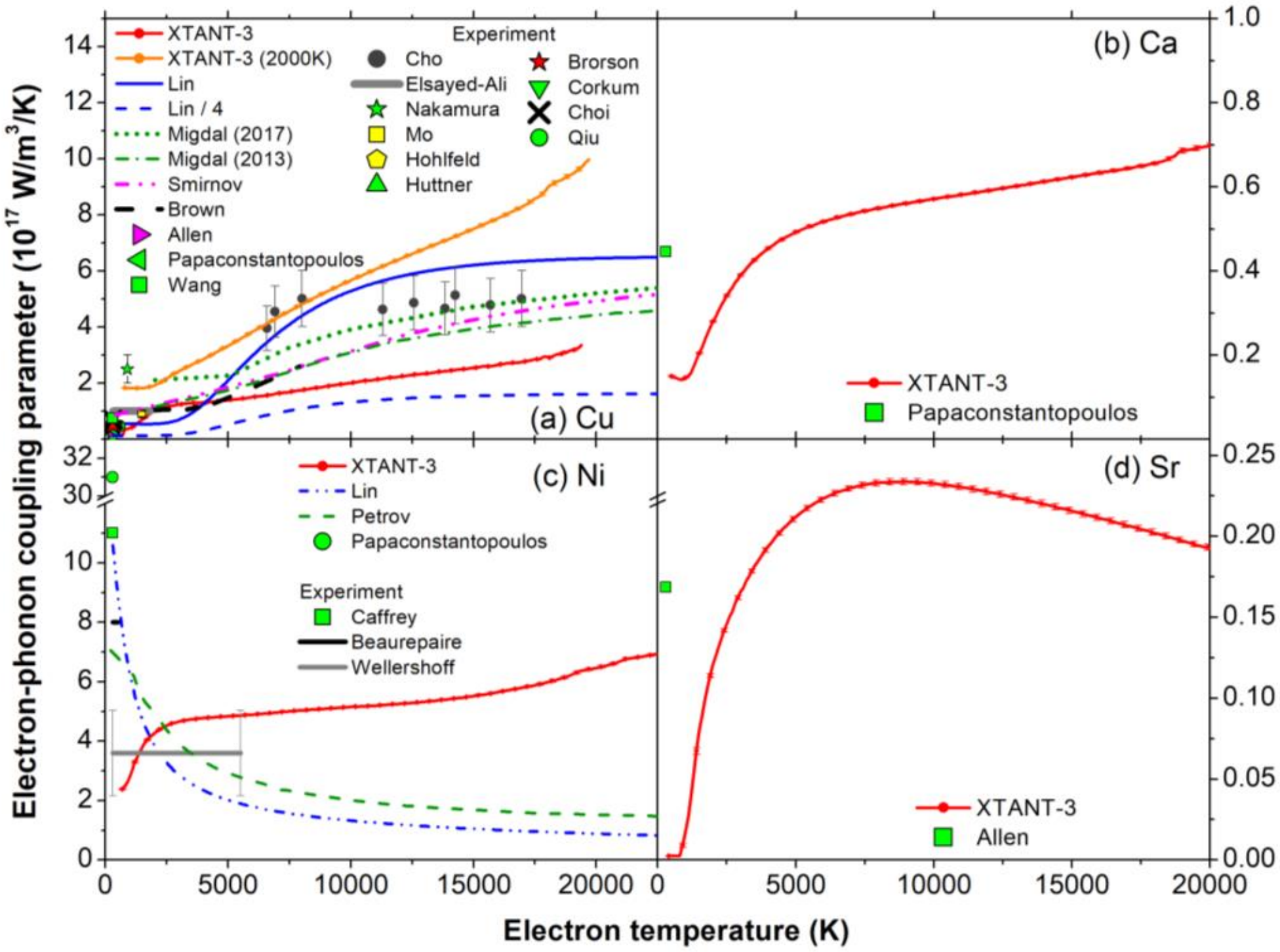

*Figure 2. Electron-phonon coupling parameter in: (a)* ***Copper****, calculated within XTANT-3 dynamical coupling approach calculated at the atomic temperatures of 300K and 2000K, compared with various theoretical and experimental data available. Experimental data include those of Hohlfeld et al.* [36]*, Qiu et al.* [47]*, Nakamura et al.* [37]*, Corkum et al.* [48]*, Elsayed-Ali et al.* [49]*, Huttner et al.* [32]*, Brorson et al.* [38]*, Mo et al.* [50]*, Choi et al.* [51]*, and Cho et al.* [52]*. Theoretical estimates are by Lin et al.* [16]*, Brown et al.* [18]*, Papaconstantopoulos* [30]*, Allen et al.* [59]*, Wang et al.* [45]*, Migdal et al.* [53] *and* [44]*, and Smirnov et al.* [43]*.*

*(b)* ***Calcium****, calculated within XTANT-3 dynamical coupling approach, compared with a theoretical estimation by Papaconstantopoulos* [30]*.*

*(c)* ***Nickel****, calculated within XTANT-3 dynamical coupling approach, compared with various theoretical and experimental data available. Experimental data include those of Caffrey et al.* [54]*, Beaurepaire et al.* [55]*, and Wellershoff et al.* [56]*. Theoretical estimates are by Lin et al.* [16]*, Petrov et al.* [17]*, and Papaconstantopoulos* [30]*.*

*(d)* ***Strontium****, calculated within XTANT-3 dynamical coupling approach, compared with the estimation by Allen et al.* [28]*.*

We also point out that Corkum's *et al.* [48] experimental data on the coupling strength in copper are obtained from multishot pulse experiments. It resulted in the coupling parameter being more than an order of magnitude lower than that from other technics employing single shot ultrashort laser pulses. The fact that different experimental techniques measure different values of coupling strength also indicates that it is sensitive to other parameters, not only the electronic temperature.

Electron-phonon coupling in calcium is presented in Figure 2b. In this case, there is no experimental data to compare to that we are aware of. A comparison with a theoretical estimation from Ref. [30] at a near-zero electronic temperature suggests that the XTANT-3 predicted drop with a decrease of $T_e$ may be too strong. As already mentioned above, that may be due to the fact that the calculations are performed for the single gamma-point in the Brillouin zone. Future experimental investigations should validate the predictions and help to benchmark the results.

Electron-phonon coupling in nickel is shown in Figure 2c. We note here that the calculations by Lin *et al*. for nickel were adjusted in the original work [16] to the experimental room-temperature data point by Caffrey *et al.* [54]. Lin's *et al.* [16] and Petrov's *et al.* [17] calculations predict a decrease of the coupling parameter with the electronic temperature increase. It is attributed by the authors of the original works to the fact that smearing of the Fermi distribution removes electrons from the energy close to the Fermi energy, which has a sharp peak in the DOS in nickel. XTANT-3 calculations predict qualitatively different behavior of the coupling strength with the increase of $T_e$: it shows a continuous increase. We attribute this increase to the nonthermal effects influencing the atomic potential and dynamics [23]. Excitation of electrons to a high temperature affects the interatomic potential as we mentioned above, which triggers atomic excitation by the mechanism known as displacive excitation of coherent phonons [57]. In turn, it leads to a stronger coupling between the electrons and the phonons. This is an inherently dynamical effect that could not be reproduced in the static calculations such as those in Refs. [16,17]. It is unclear if accounting for the modification of the phonon spectrum at elevated $T_e$ in these static calculations would fully account for such an effect, or there may be some synergetic effect at play, but this problem is beyond the scope of the current work.

Figure 2d demonstrates the electron-phonon coupling in strontium calculated with the XTAN-3 code. As we noted in the model section, the simulations are performed for the single gamma point. In the case of strontium, it produces a small band gap of ~0.15 eV, because the dispersion curves are supposed to meet at the L point in the Brillouin zone [58], which is not sufficiently sampled for the supercells of our size (500 atoms in this case). Thus, one can expect a reliable description of the coupling parameter only at electronic temperatures above the value of this artificial band gap, which corresponds to $T_e$ ~ 2-3 kK. Thus, the curve bending downwards with the decrease of the electronic temperature in Figure 2d may rather be an artifact. For practical purposes, one may use the low-temperature point calculated by Allen *et al*. [28] and interpolate to our calculated data.

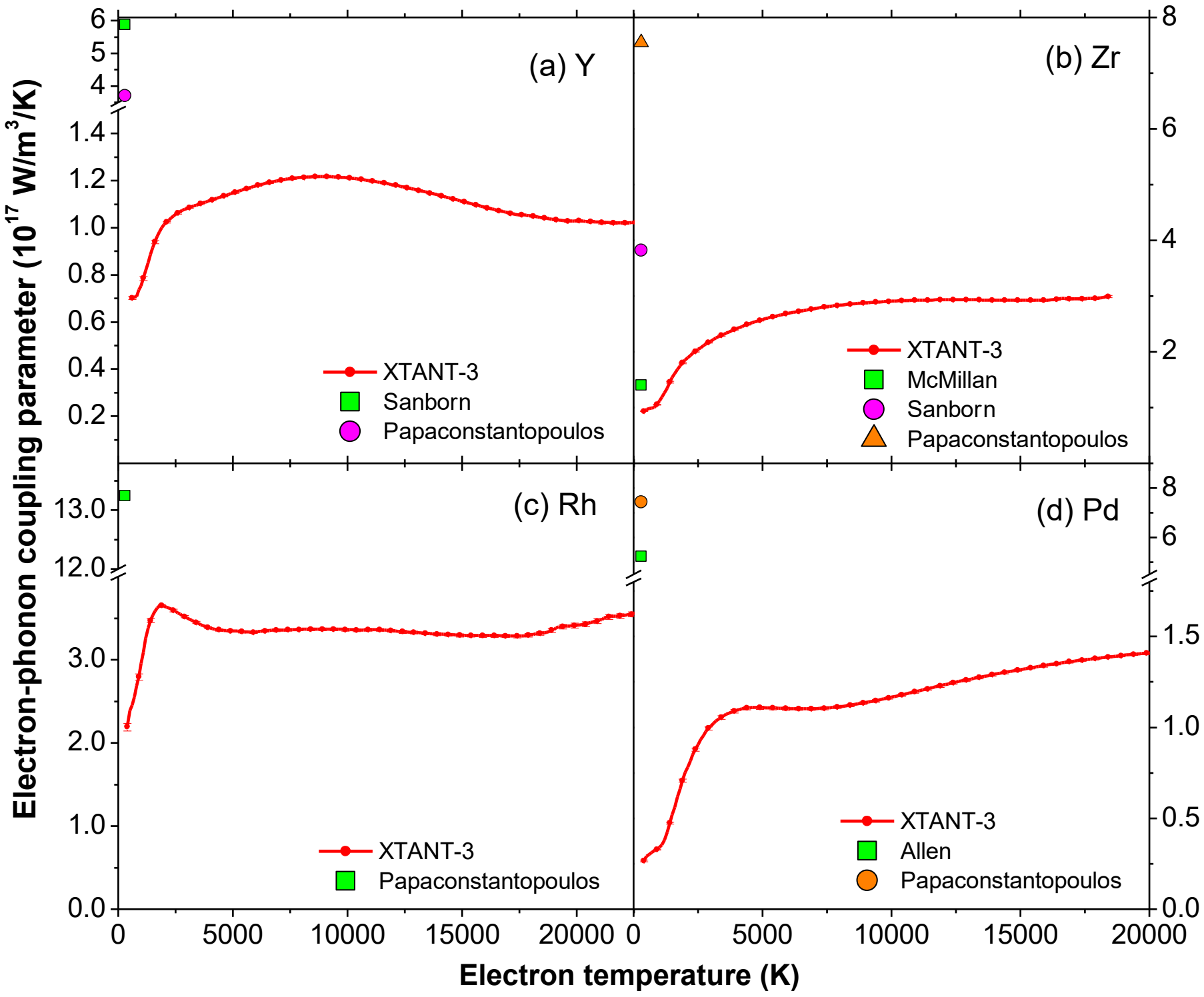

*Figure 3. Electron-phonon coupling parameter in: (a)* ***Yttrium*** *calculated within XTANT-3 dynamical coupling approach, compared with theoretical estimates by Papaconstantopoulos* [30] *and Sanborn et al.* [31].
*(b)* ***Zirconium*** *calculated within XTANT-3 dynamical coupling approach, compared with theoretical estimates McMillan* [29], *Sanborn et al.* [31], *and Papaconstantopoulos* [30].
*(c)* ***Rhodium*** *calculated within XTANT-3 dynamical coupling approach, compared with the theoretical estimate by Papaconstantopoulos* [30].
*(d)* ***Palladium*** *calculated within XTANT-3 dynamical coupling approach, compared with the theoretical estimates by Allen et al.* [59], *and Papaconstantopoulos* [30].

The electron-phonon coupling parameter in yttrium is displayed in Figure 3a. Our calculations are lower than other theoretical estimates, however, those also differ by a factor of 1.5 between themselves. In the absence of experimental data, we cannot judge on which data should be used in future modelling.

The electron-phonon coupling parameter in zirconium is shown in Figure 3b. Here, our calculations are close to the semi-empirical results by McMillan [29] at low temperatures, but there is no experimental data to compare with.

Figure 3c shows the electron-phonon coupling parameter in rhodium. Again, another theoretical result for near-zero temperature coupling [30], estimated from the electron-phonon mass enhancement, is significantly higher than our calculations, but it could not be cross checked against an experiment.

A similar situation occurs in palladium, shown in Figure 3d. Here as well the XTANT-3 calculations show an electron-phonon coupling parameter significantly lower than other estimations. This should not concern us too much, since, as we will see throughout this paper, those estimations do not always agree well with the experimental data when available, even at low temperatures. They are merely shown here for completeness and to illustrate how different the results may be depending on a particular model used.

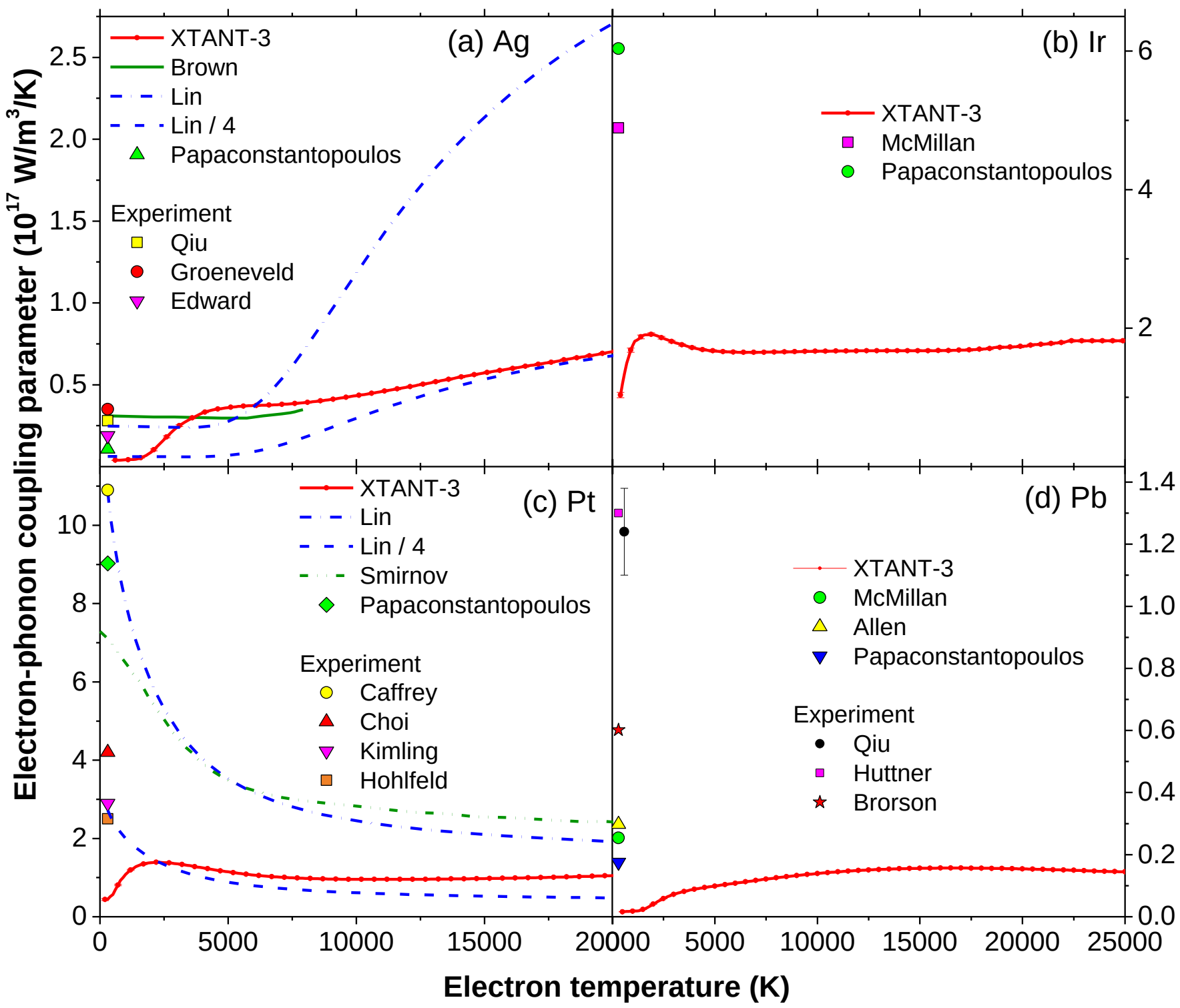

*Figure 4. Electron-phonon coupling parameter in: (a)* ***Silver****, calculated within XTANT-3 dynamical coupling approach, compared with various theoretical and experimental data available. Experimental data include those of Qiu et al.* [47]*, Groeneveld et al.* [60]*, and Edward et al.* [61]*.Theoretical estimates are by Lin et al.* [16] *(and the same data rescaled by the factor of 1/4), Brown et al.* [18]*, and Papaconstantopoulos* [30].

*(b)* ***Iridium****, calculated within XTANT-3 dynamical coupling approach, compared with the theoretical estimates by McMillan* [29]*, and Papaconstantopoulos* [30].

*(c)* ***Platinum****, calculated within XTANT-3 dynamical coupling approach, compared with various theoretical and experimental data available. Experimental data include those of Hohlfeld et al.* [62]*, Kimling et al.* [63]*, Choi et al.* [51]*, and Caffrey et al.* [54]*. Theoretical estimates are by Lin et al.* [16] *(and the same data rescaled by 1/4), and Papaconstantopoulos* [30].

*(d)* ***Lead****, calculated within XTANT-3 dynamical coupling approach, compared with various theoretical and experimental data available. Experimental data include those of Qiu et al.* [47]*, Huttner et al.* [32]*, and Brorson et al.* [38]*. Theoretical estimates are by McMillan* [29]*, Allen et al.* [28]*, and Papaconstantopoulos* [30].

Calculations for silver, shown in Figure 4a, are below the predictions by Lin *et al.* [16], however, they are close to the more recent *ab-initio* calculations by Brown *et al.* [18] which show a slower increase of the coupling with an increasing electronic temperature. At higher electronic temperatures, XTANT-3 calculations surprisingly match those by Lin *et al.* rescaled by a factor of 1/4. A comparison at low electronic temperatures indicates that in XTANT-3 a drop of the coupling strength below ~2500 K is probably an artifact. Experimental data at near-room-temperature $T_e$ differ by a factor of two among themselves, and do not allow to discriminate among the theoretical approaches. High-electron-temperature experiments would provide a clearer ground for validating models, since there the different models' results diverge more significantly.

Electron-phonon coupling in iridium is shown in Figure 4b. A comparison with the near-zero theoretical estimations shows lower values in our results.

The electron-phonon coupling parameter in platinum is displayed in Figure 4c. We emphasize that the calculations by Lin *et al.* [16] for platinum were adjusted to the experimental room temperature data point of Caffrey *et al.* [54]. Alternatively, fitting to the data by Kimling *et al.* [63] or Hohlfeld *et al.* [36] results into a lower lying curve (marked as "Lin/4" in Figure 4c since it exactly corresponds to the rescaling of the original data by a factor of 1/4), closer to the prediction of XTANT-3. The behavior of the curves is similar to that in nickel discussed above. The characteristic increase of the coupling strength visible in XTANT-3 calculations but absent in those by Lin's *et al.* is again attributed to the interplay with the nonthermal effects.

Figure 4d shows the coupling strength in lead. Here, our results are relatively close to the three theoretical low-temperature estimations from Refs. [28–30]. All of the theoretical estimations, including ours, lie significantly lower than the available experimental points at low $T_e$ (although the experimental data differ among themselves by the factor of two [32,38,47]). This may indicate an existence of some effects that are not captured in any of the theoretical approaches, and seem to deserve future dedicated studies.

## III.2. Hcp metals

Now we proceed to the hcp elemental metals: Mg, Sc, Ti, Co, Zn, Tc, Ru, Cd, Hf, Re, and Os. For these calculations, we used supercells composed of 5x3x4 conventional orthogonal unit cell (240 atoms) for Sc, Co, Zn, Tc, Cd, Hf, Re, Os; 5x3x5 unit cells (300 atoms) for Mg; 6x3x3 unit cells (216 atoms) for Cd; and 5x5x5 unit cells (500 atoms) for Ti and Ru.

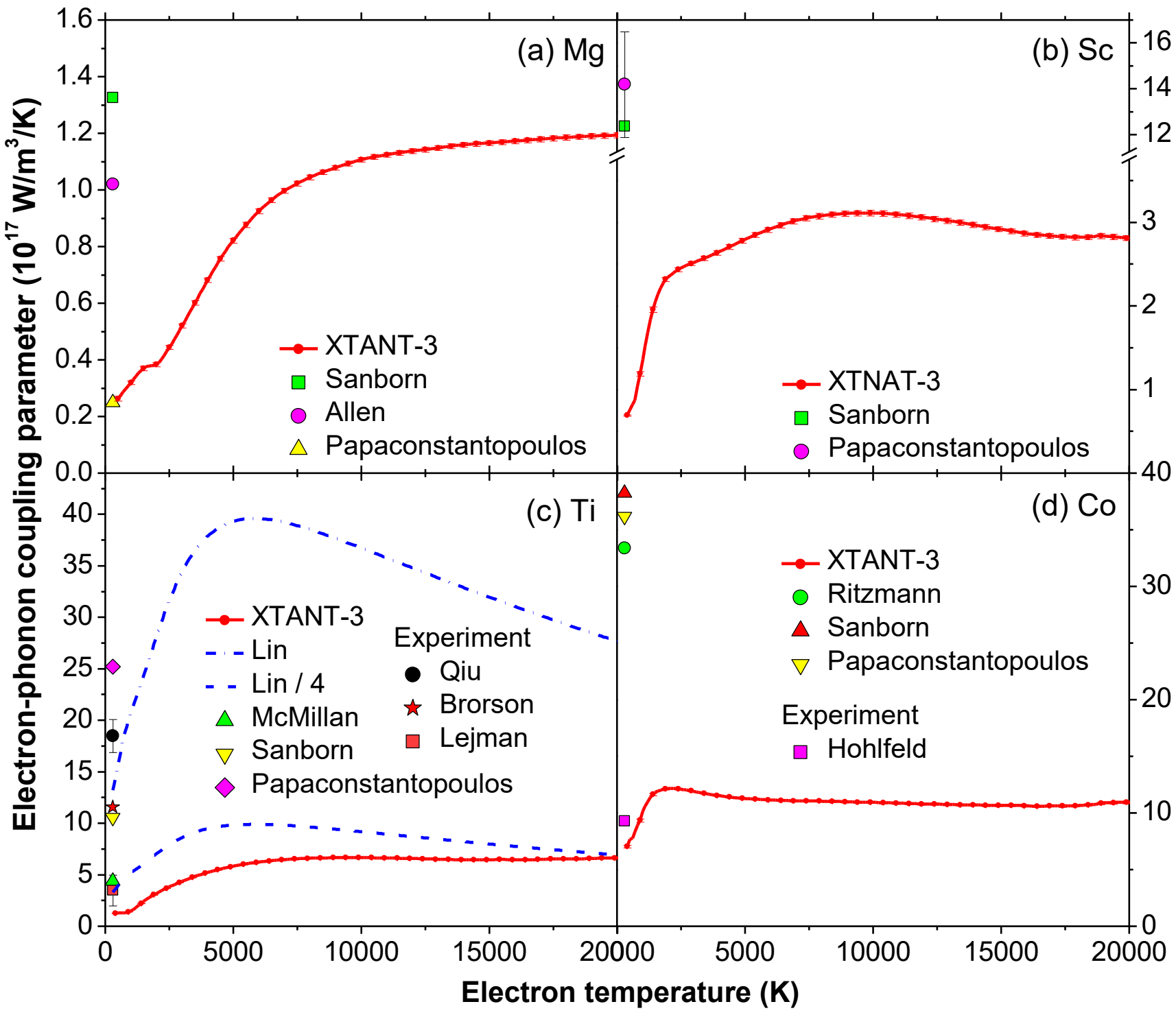

*Figure 5. Electron-phonon coupling parameter in: (a)* ***Magnesium****, calculated within XTANT-3 dynamical coupling approach, compared with the theoretical estimates by Allen et al.* [28]*, Sanborn et al.* [31]*, and Papaconstantopoulos* [30]*.*
*(b)* ***Scandium****, calculated within XTANT-3 dynamical coupling approach, compared with the theoretical estimates by Sanborn et al.* [31]*, and Papaconstantopoulos* [30]*.*
*(c)* ***Titanium****, calculated within XTANT-3 dynamical coupling approach, compared with various theoretical and experimental data available. Experimental data include those of Qiu et al.* [47]*, Brorson et al.* [38]*, and Lejman et al.* [64]*. Theoretical estimates are by Lin et al.* [16] *(and the same data rescaled by 1/4), McMillan* [29]*, Sanborn et al.* [31]*, and Papaconstantopoulos* [30]*.*
*(d)* ***Cobalt****, calculated within XTANT-3 dynamical coupling approach, compared with experimental data by Hohlfeld et al.* [36]*, and theoretical estimates by Sanborn et al.* [31]*, Ritzmann et al.* [65]*, and Papaconstantopoulos* [30]*.*

Figure 5a demonstrates an electron-phonon coupling strength in magnesium calculated with XTANT-3 and compared with other theories. Our low-temperature results lie close to the ones by Papaconstantopoulos [30], but there is no experimental data to compare to so far.

Figure 5b shows a calculated coupling parameter in scandium. Here, our calculations are lower than other theoretical estimations, while again no experimental data are available to validate the models.

In titanium, shown in Figure 5c, Lin *et al.* calculations [16] were adjusted to the experimental point by Brorson *et al*. [38] setting the room temperature value of the coupling strength. We note that if, e.g., McMillan's [29] value of the electron-phonon mass enhancement were used, it would result into a lower lying curve, rescaled again by the same factor of 1/4 (see Figure 5c). Such a rescaled curve overlaps closer with our XTANT-3 calculations, showing a qualitative similarity of the results. Experimental data here differ significantly even at the near-room temperature, and no data are available at the high electronic temperatures.

In cobalt, the electron-phonon coupling strength shown in Figure 5d demonstrates a case when our calculations are very close to the available experimental datum from Ref. [36], whereas all other models overestimate it by about a factor of 4. This fact supports our claim that in the previously discussed cases when other theories produced results high above ours, those discrepancies did not tell much on the validity of any model involved – they all ultimately require a validation against experiments.

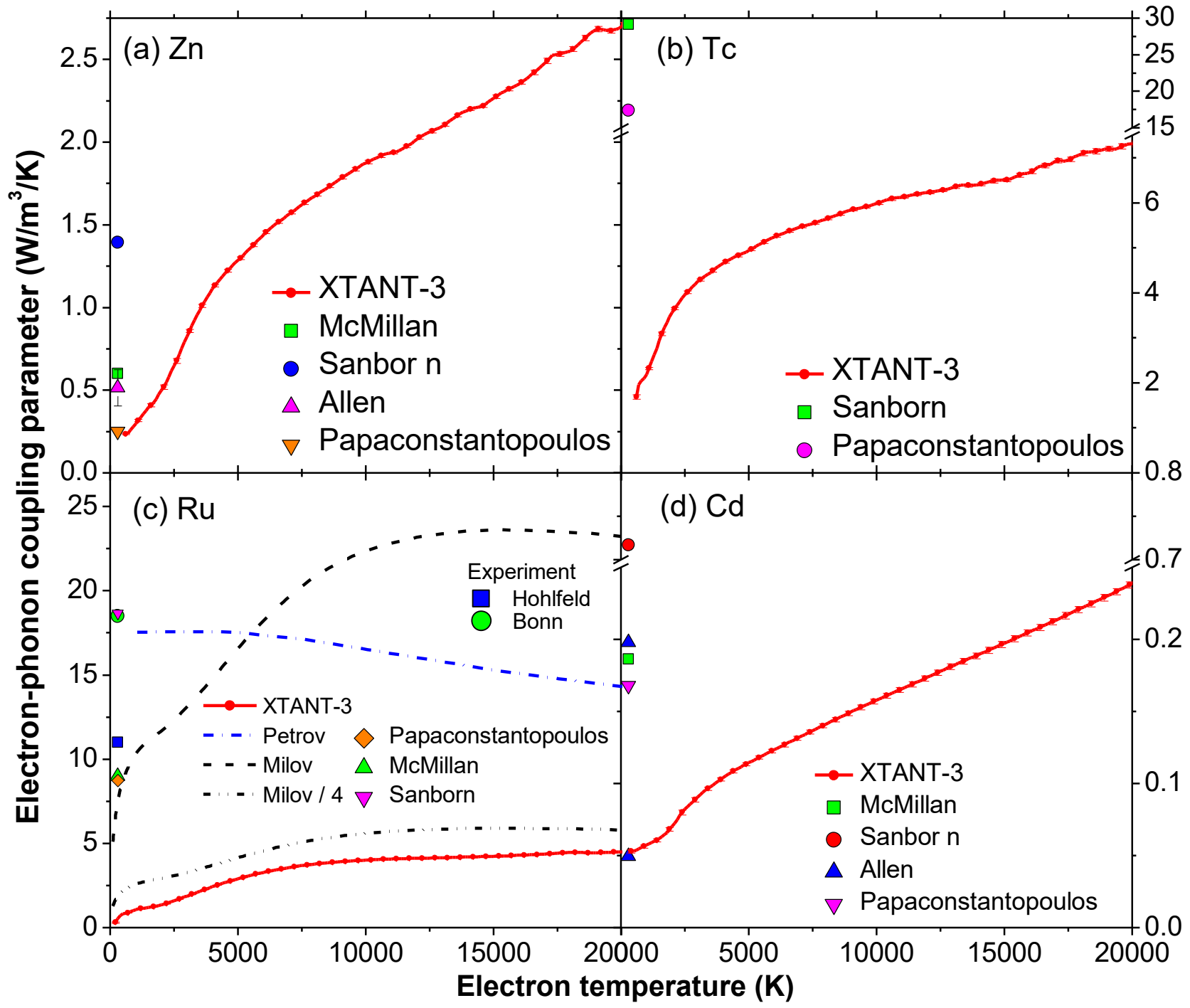

*Figure 6. Electron-phonon coupling parameter in: (a)* ***Zinc****, calculated within XTANT-3 dynamical coupling approach, compared with the theoretical estimates by Allen et al.* [28], *McMillan* [29], *Sanborn et al.* [31], *and Papaconstantopoulos* [30].
*(b)* ***Technetium****, calculated within XTANT-3 dynamical coupling approach, compared with the theoretical estimates by Sanborn et al.* [31], *and Papaconstantopoulos* [30].
*(c)* ***Ruthenium*** *calculated within XTANT-3 dynamical coupling approach, compared with various theoretical and experimental data available. Experimental data include those of Hohlfeld et al.* [36], *and Bonn et al.* [66]. *Theoretical estimates are by McMillan* [29], *Sanborn et al.* [31], *Papaconstantopoulos* [30], *Petrov et al.* [67], *and Milov et al.* [68] *(and the same data rescaled by the factor of 1/4).*
*(d)* ***Cadmium*** *calculated within XTANT-3 dynamical coupling approach, compared with the theoretical estimates by Allen et al.* [28], *McMillan* [29], *Sanborn et al.* [31], *and Papaconstantopoulos* [30].

XTANT-3 calculations for zinc are shown in Figure 6a. Here, other theoretical models disagree among themselves by a factor of five, and so they do with our results.

Figure 6b shows the electron-phonon coupling in technetium, where again our results demonstrate much lower coupling strength than other theories at near-zero electron temperature.

Figure 6c demonstrates the coupling parameter in ruthenium. Note that in our previous work on the electron-phonon coupling in ruthenium in Refs. [68,69], there was an error that resulted in overestimation of the increase of the coupling parameter with increase of the electronic temperature. The corrected curve is plotted here in Figure 6c (marked as "Milov"). This calculations rely on the formalism of the Boltzmann collision integral, similar to that of Allen *et al*. [70], extended to the finite temperature regime according to Muller and Rethfeld [35]. The method is analogous to the one used to obtain the curve "Gorbunov 3" for Al discussed above. As was pointed out in Ref. [24], a usage of the simplistic jelly-like model for the matrix element is sensitive to the choice of the screening parameter, and may overestimate the coupling strength. We thus rescale the calculations by Milov *et al*. [68,69] by the same factor of 1/4, as was done above for the results of Lin *et al*. in other metals, and find a good agreement with the XTANT-3 calculations (see Figure 6c, a comparison with the curve marked as "Milov/4"). However, such results do not match the experimental points at the room temperature. A recent calculation by Petrov *et al*. [67] based on DFT calculations with different effective masses for *s*- and *d*-electrons exhibits a qualitatively different behavior of the coupling parameter compared to other calculations, namely a slow decrease with increasing electron temperature. This calculation agrees with the experimental data point by Bonn *et al*. [66] at the room temperature, but disagrees with the other one by Hohlfeld *et al*. [62]. The experimental data points disagree among themselves by almost a factor of two. A spread of experimental data indicates the need for more experiments to draw a reliable conclusion about accuracy of various calculation schemes.

The electron-phonon coupling parameter in cadmium is plotted in Figure 6d. Here, our results are close to one of Allen *et al*.'s estimations [28]. The work [28] showed various calculations made with different empirical pseudopotentials and phonon spectra, which resulted in noticeably different electron-phonon mass enhancement parameters. Cadmium is a case where different models predict very different coupling parameters.

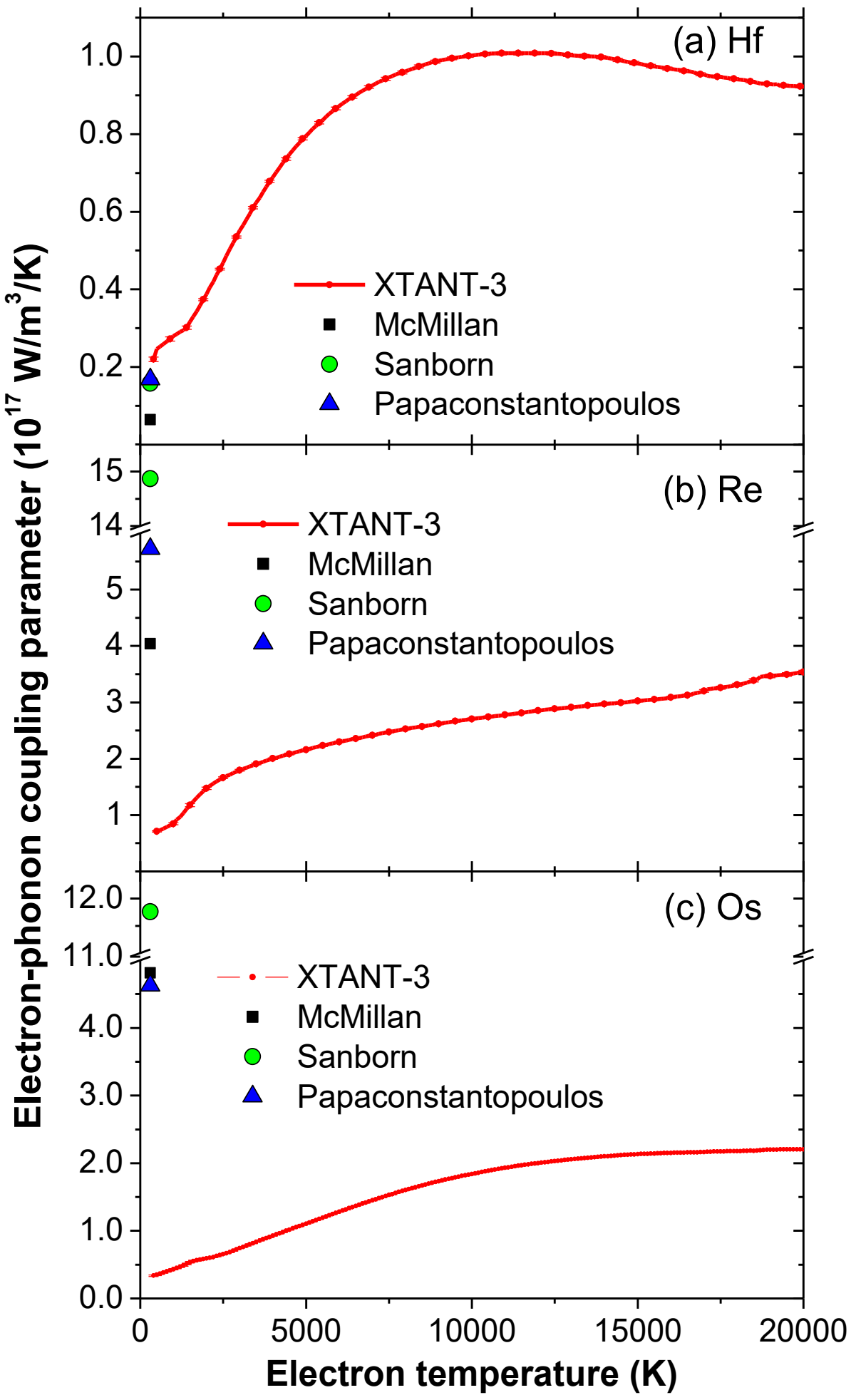

*Figure 7. Electron-phonon coupling parameter in: (a)* ***Hafnium****, calculated within XTANT-3 dynamical coupling approach, , compared with the theoretical estimates by McMillan* [29]*, Sanborn et al.* [31]*, and Papaconstantopoulos* [30]*.*
*(b)* ***Rhenium****, calculated within XTANT-3 dynamical coupling approach, compared with the theoretical estimates by McMillan* [29]*, Sanborn et al.* [31]*, and Papaconstantopoulos* [30]*.*
*(c)* ***Osmium****, calculated within XTANT-3 dynamical coupling approach, compared with the theoretical estimates by McMillan* [29]*, Sanborn et al.* [31]*, and Papaconstantopoulos* [30]*.*

The coupling parameter in hafnium, shown in Figure 7a, agrees reasonably well with other models estimations in low-electron-temperature regime, except for the estimation by McMillan [29].

In rhenium, the electron-phonon coupling parameters shown in Figure 7b exhibit large scattering among the various theoretical models at the near-zero electron temperature. No experimental data are available yet to validate the models.

A similar situation is with the coupling parameter in osmium in Figure 7c, where the models disagree among themselves.

### III.3. Bcc metals

The following examples of the bcc metals are studied in this section: V, Cr, Fe, Nb, Mo, Ba, Ta, and W. The supercell is composed of 5x5x5 conventional orthogonal unit cells (250 atoms) in all the cases.

The calculation of the electron-phonon coupling with XTANT-3 code in vanadium is plotted in Figure 8a. The theoretical low-temperature estimations with other models disagree among themselves and with our results. The experimental data at the near-room temperature also differ among various authors, but all the cases are higher than the results of our calculations. One reason for this is, of course, that our calculations at low electronic temperatures may underestimate the data.

We see that from $T_e$~6000 K down to $T_e$~2000 K the coupling parameters increases, and may be further extrapolated to rise closer to experimental data at the room temperature.

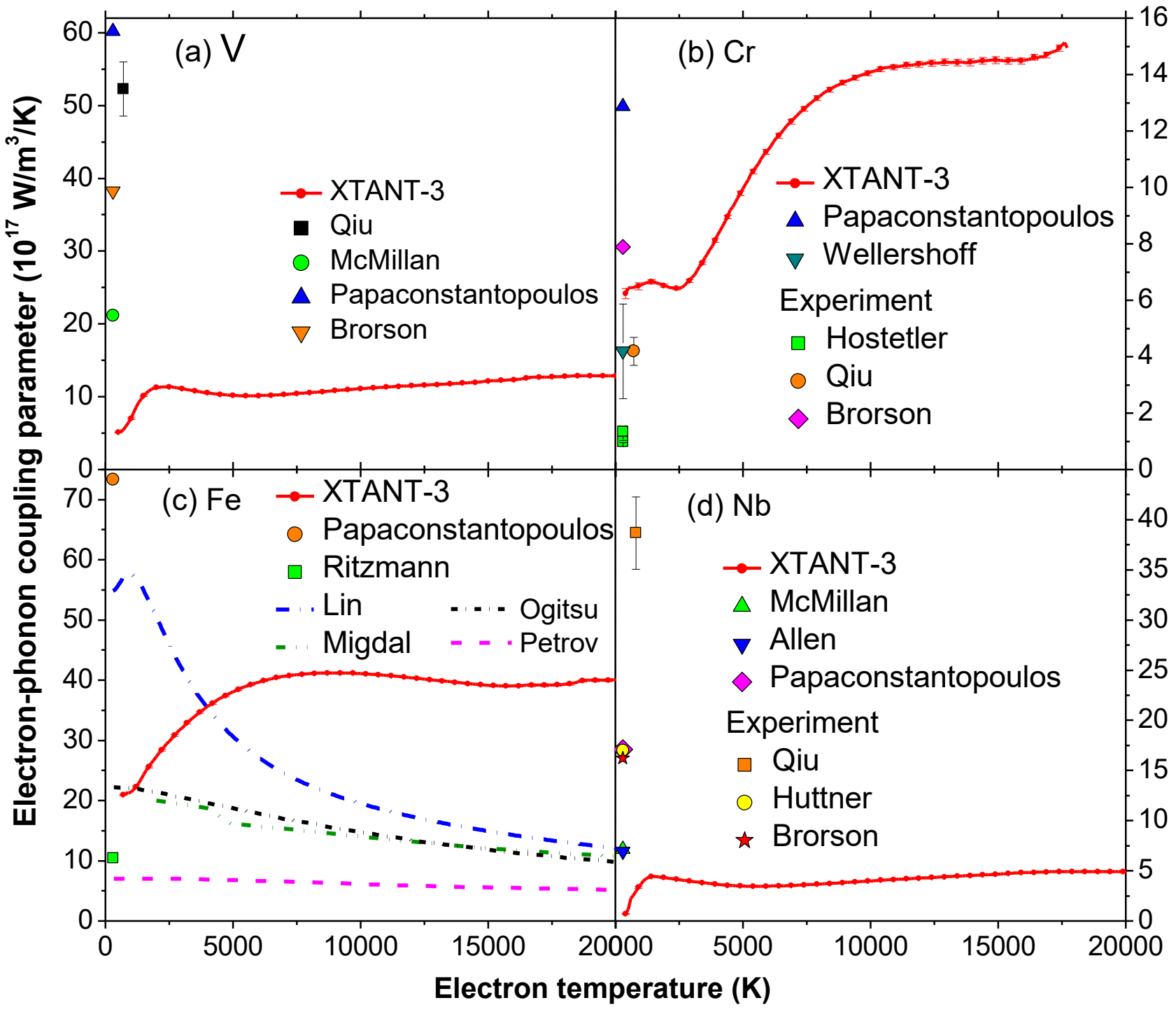

*Figure 8. Electron-phonon coupling parameter in: (a)* ***Vanadium****, calculated within XTANT-3 dynamical coupling approach, compared with various theoretical estimates by Qiu et al.* [47]*, Brorson et al.* [38]*, McMillan* [29]*, and Papaconstantopoulos* [30]*.*
*(b)* ***Chromium****, calculated within XTANT-3 dynamical coupling approach, compared with various theoretical and experimental data available. Experimental data include those of Qiu et al.* [47]*, Hostetler et al.* [33]*, and Brorson et al.* [38]*. Theoretical estimates are by Wellershoff et al.* [56]*, and Papaconstantopoulos* [30]*.*
*(c)* ***Iron****, calculated within XTANT-3 dynamical coupling approach, compared with theoretical estimates by Ritzmann et al.* [65]*, Papaconstantopoulos* [30]*, Lin et al.* [16]*, Petrov et al.* [17]*, Migdal et al.* [53]*, and Ogitsu et al.* [71]*.*
*(d)* ***Niobium****, calculated within XTANT-3 dynamical coupling approach, compared with various theoretical and experimental data available. Experimental data include those of Qiu et al.* [47]*, Huttner et al.* [32]*, and Brorson et al.* [38]*. Theoretical estimates are by McMillan* [29]*, Allen et al.* [59]*, and Papaconstantopoulos* [30]*.*

The electron-phonon coupling parameter in chromium is shown in Figure 8b. Here our results are close to some of the experimental data at room temperature, however, the experimental data scatter significantly, making it hard to draw a conclusion.

Our calculated coupling parameter in iron is shown in Figure 8c together with various theoretical estimations. The data found in the literature differ by an order of magnitude, precluding us from a meaningful comparison. Qualitatively, the situation is similar to that in nickel and platinum: XTANT-3 predicts an increase of the coupling strength with increasing $T_e$, whereas other predictions show only a decrease. The calculations by Migdal *et al.* [53] and Ogitsu *et al.* [71] used the same methodology and thus agree between themselves.

Figure 8d shows the electron-phonon coupling parameter in niobium. Here, experimental data at the near-room temperature differ among themselves by a factor of ~2.5. All theoretical estimations lie lower than the experimental data. Similar to the case of Pb above, it indicates that niobium requires a dedicated research, preferably in both directions, experimental and theoretical.

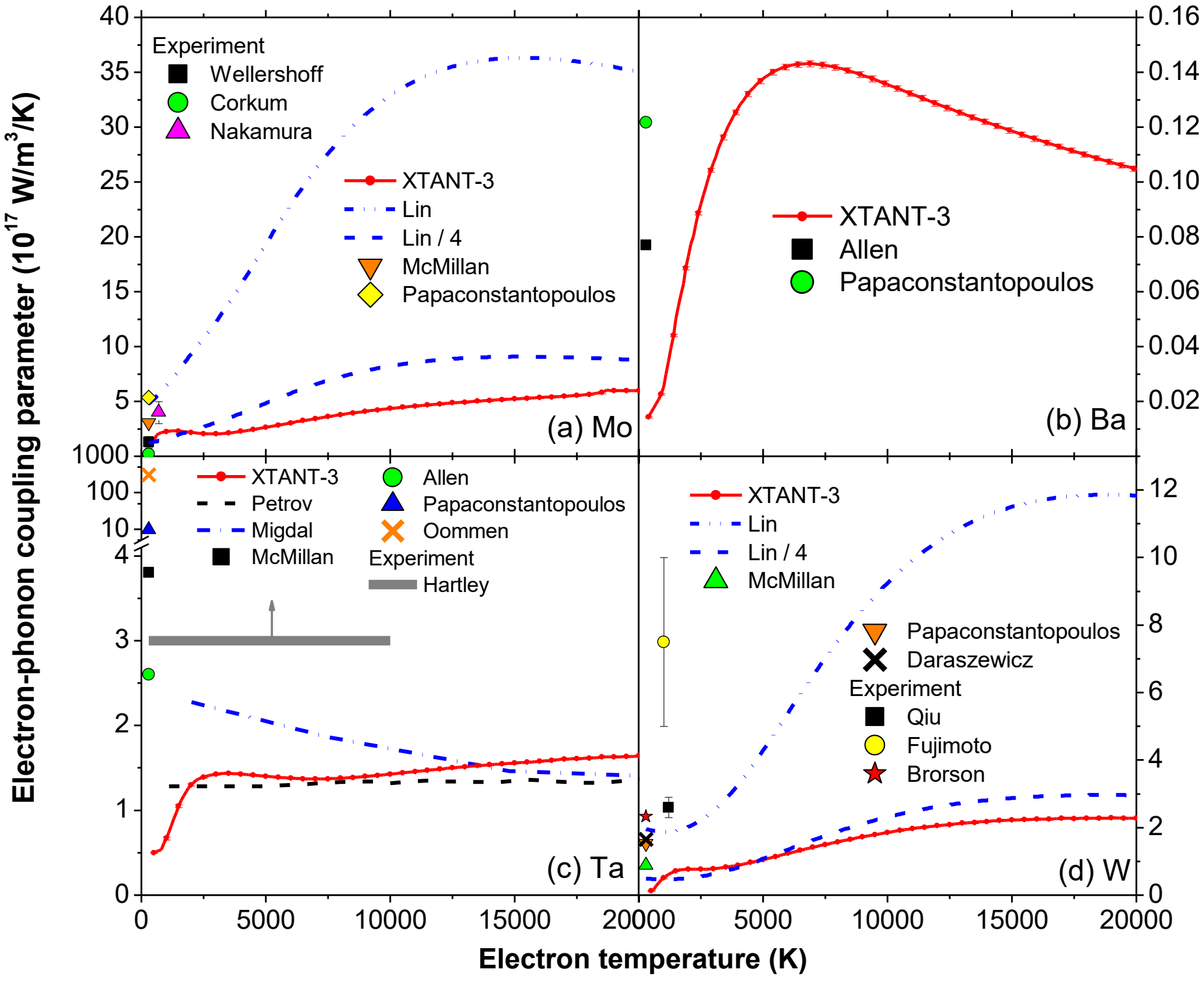

*Figure 9. Electron-phonon coupling parameter in: (a)* ***Molybdenum****, calculated within XTANT-3 dynamical coupling approach, compared with various theoretical and experimental data available. Experimental data include those of Nakamura et al.* [37]*, Corkum et al.* [48]*, and Wellershoff et al.* [56]*. Theoretical estimates are by McMillan* [29]*, Papaconstantopoulos* [30]*, Lin et al.* [16] *(and the same data rescaled by 1/4).*

*(b)* ***Barium****, calculated within XTANT-3 dynamical coupling approach, compared with other theoretical estimates Allen et al.* [28]*, and Papaconstantopoulos* [30].

*(c)* ***Tantalum****, calculated within XTANT-3 dynamical coupling approach, compared with the experimental estimation by Hartley* et al. [72] (an arrow indicates that it is a lower boundary estimate), and the*theoretical models by Migdal et al.* [53]*, Petrov et al.* [17]*, McMillan* [29]*, Allen et al.* [59]*, Papaconstantopoulos* [30]*, and Oommen et al.* [73].

*(d)* ***Tungsten****, calculated within XTANT-3 dynamical coupling approach, compared with various theoretical and experimental data available. Experimental data include those of Qiu et al.* [47]*, Fujimoto et al.* [74]*, and Brorson et al.* [38]*. Theoretical estimates are by McMillan* [29]*, Papaconstantopoulos* [30]*, Daraszewicz et al.* [75]*, Lin et al.* [16] *(and the same data rescaled by 1/4).*

The various results on the electron-phonon coupling parameter in molybdenum are collected and plotted in Figure 9a. Here, the results by Lin *et al.* [16] show a typical increase to much higher values with the increase of $T_e$ than our results. If rescaled in such a way as to agree at low-electronic-temperature limit with the experiment by Wellershoff *et al.* [56], the same way as was done for aluminum and other materials, it would go much closer to the XTANT-3 results. Curiously enough, such a rescaling again requires the factor of 1/4. Various experimental data differ significantly among themselves. The data by Corkum *et al.* [48] are again much lower than all the other data, since they are obtained in the multishot experiments.

The electron-phonon coupling parameter in barium is shown in Figure 9b. Theoretical estimations by other authors are not far off XTANT-3 calculations.

Figure 9c shows the coupling strength in tantalum compared with various theoretical and empirical approaches and an experimental estimation by Hartley *et al.* [72]. The calculations by Petrov *et al.* [17] and by Migdal *et al.* [53] at high electronic temperatures agree well with our results. Near-zero temperature theoretical estimates, however, scatter in a wide range. One of the most recent estimates suggests an extremely high value of $3.1 \times 10^{19}$ W/(m$^3$K) [73], which strikes us as rather unrealistic. The experimental estimation from Ref. [72] is for the lower boundary on the coupling parameter in the warm dense state in tantalum created by 10-ps-long laser pulse, thus it may not correspond to the conditions we and other authors used in the simulations. Nonetheless, our obtained estimate agrees with most of the models within about a factor of two.

Figure 9d displays the electron-phonon coupling parameter in tungsten. The XTANT-3 calculations are lower than those by Lin *et al*. [16], which is not surprising having in mind other cases analyzed above. And again, rescaled by a factor of 1/4, Lin's data are very close to ours, indicating a qualitative similarity. Experimental data here at the near-room temperature are closer to Lin's *et al*., however, different authors reported the data differing by nearly an order of magnitude.

### III.4. Diamond cubic lattice

We analyzed the diamond cubic lattice metal Sn, and two semiconductors with relatively small band gaps: Si and Ge. Here, the supercell is composed of 3x3x3 conventional orthogonal unit cells (216 atoms).

The electron-phonon coupling in tin is shown in Figure 10a. The other theoretical estimates at the low electronic temperature are much higher than the XTANT-3 calculations. However, the recent experiment by Waldecker et al. [76] shows that an average coupling parameter at the corresponding temperatures is closer to our results. Transient atomic diffraction was measured in the experiment, and these data were used to reconstruct the atomic temperature evolution after an ultrafast laser pulse irradiation. The average coupling parameter was then obtained by fitting the two-temperature model to the evolution of the atomic temperature data. The constant value, corresponding to the best fit [76], within the corresponding electron temperature range is plotted in Figure 10a.

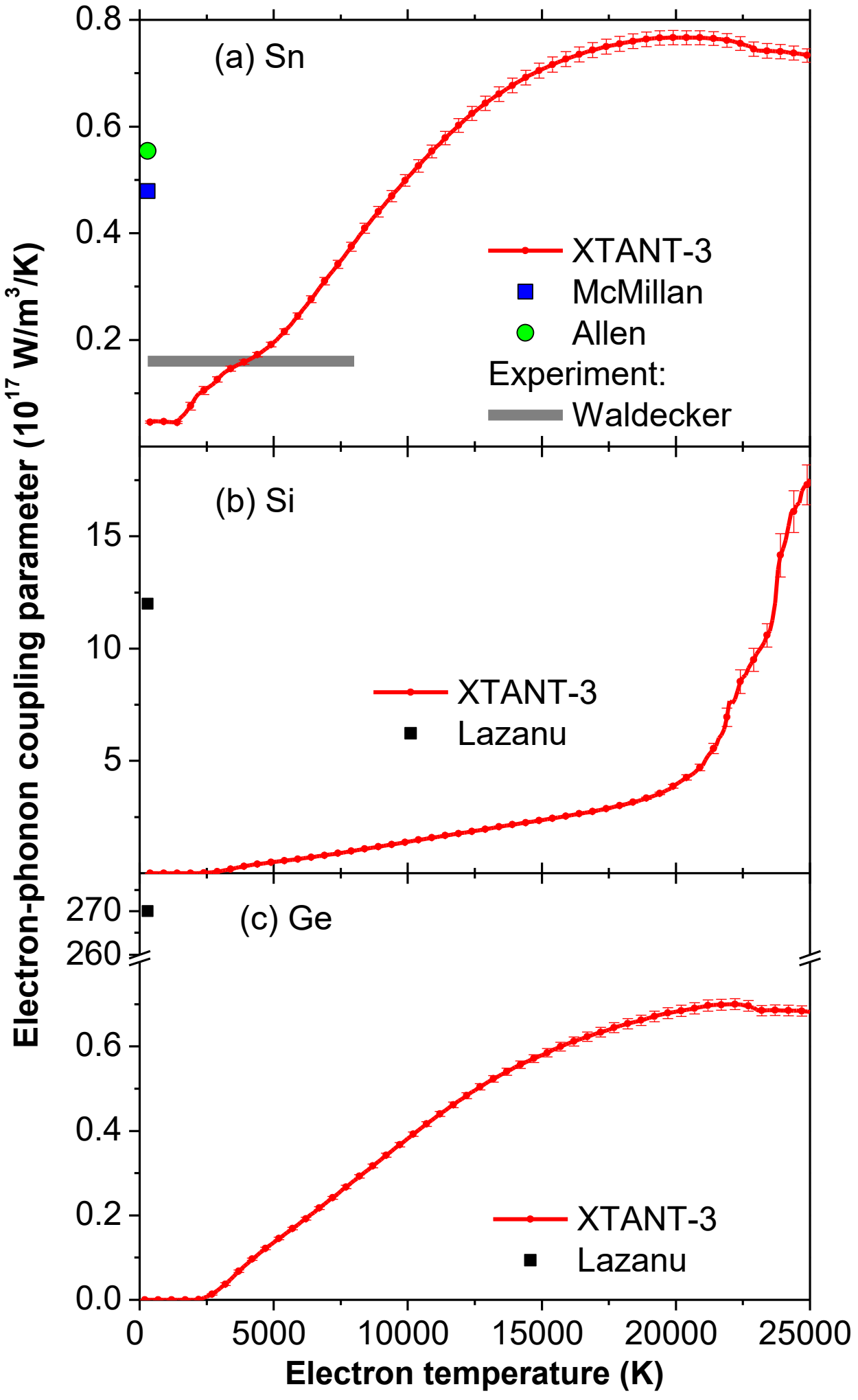

*Figure 10. Electron-phonon coupling parameter in: (a)* ***Tin****, calculated within XTANT-3 dynamical coupling approach, compared with the experimental data from Waldecker* [76]*, and the theoretical estimates by McMillan* [29]*, Allen et al.* [28]*, and Papaconstantopoulos* [30].
*(b)* ***Silicon****, calculated within XTANT-3 dynamical coupling approach, compared with a theoretical value from Lazanu et al.* [77].
*(c)* ***Germanium*** *calculated within XTANT-3 dynamical coupling approach, compared with a theoretical value from Lazanu et al.* [77].

The electron-phonon coupling in silicon and germanium are shown in Figure 10b and c, correspondingly. Here we do not have reliable data to cross check against. There is almost no electrons in the conduction band of a semiconducting material at low temperatures, and thus no coupling to phonons. The values from Ref. [77] do not seem to be physically meaningful, as they are merely fitting parameters in the TTM used for description of fast particles tracks, so they are shown here for illustrative purposes only.

We see in both Figure 10b and c that below ~2500 K the coupling is near zero due to a lack of electrons that could overcome the material band gap for an efficient coupling to the atomic motion. When the electronic temperature is sufficiently high to overcome the band gap by thermal means, the coupling strength linearly increases up to high temperatures of $T_e$~17-19 kK in Si, after which the behavior suddenly changes demonstrating a much faster increase. This temperature corresponds to the onset of ultrafast nonthermal melting, which drastically alters the coupling too, as was discussed in detail in Refs. [21,23]. In germanium, the linear increase of the coupling is until $T_e$~12-13 kK, and it starts to saturate after that. At such temperatures, the coupling parameter becomes close to the values in metals, nearly ~$10^{17}$ W/($m^3$K).

We emphasize that the calculations for Si and Ge are performed assuming that the valence and conduction band electrons adhere to the common Fermi distribution, i.e., have the same temperature and chemical potential. It might not be the case in ultrashort laser pulse experiments. Thus, the presented data should rather be considered as a benchmark, against which other models can be cross checked. To be applicable to ultrafast laser experiments, realistic models may require a separate nonequilibrium treatment of valence and conduction bands [78]. Another important effect that must be considered in modelling covalent semiconductors is nonthermal melting, which may trigger an ultrafast phase transition in a material upon high electronic excitation before any significant electron-phonon coupling would heat up the atomic system [11].

### III.5. Other materials

We modeled alpha (A12) phase of manganese with the supercell composed of 2x2x2 conventional unit cells (464 atoms). The calculated electron-phonon coupling parameter is shown in Figure 11a. The only available point to compare here is the theoretical near-zero-temperature estimation from [30], which is significantly higher than our values at low temperatures.

Alpha-Ga (A11) was modelled within a supercell composed of 4x2x4 conventional unit cells (256 atoms). The coupling parameter in gallium is presented in Figure 11b. XTANT-3 predicts a strong increase of the coupling parameter with increase of $T_e$, but the low-temperature limit is significantly below the theoretical estimates by McMillan [29] and Papaconstantopoulos [30].

We modeled Fct (A6) structure of indium with the supercell composed of 5x5x1 conventional unit cells (225 atoms). The calculated coupling factor is shown in Figure 11c. In the low-electron-temperature limit, our results approach the estimation by McMillan [29].

To model trigonal tellurium and selenium we used a supercell composed of 4x4x4 hexagonal unit cells (192 atoms). The NRL tight binding parameterization for these materials is taken from Ref. [79]. Figure 11d displays the electron-phonon coupling in tellurium. The near-zero electron temperature value of the coupling from [30] is 9.7x$10^{13}$ W/($m^3$K), shown there for comparison, which is much lower than in any other metal. Our results produce a low-temperature limit on the order of 2x$10^{16}$ W/($m^3$K), which is closer to other metals.

The coupling parameter in selenium is presented in Figure 11e. There is no experimental or theoretical data to compare with, thus we hope our calculations shown here may serve for benchmarking in future works.

Figure 11f shows the electron-phonon coupling parameter in graphite with AB stacking, calculated for 216 atoms in the supercell. Experimental data from Ref. [80] suggested an inhibited coupling in graphite, ~0.6x$10^{16}$ W/($m^3$K). However, this estimate was done in an experiment with a 100 ps long proton bunch, and probed at over 200 ps after the pump – a different method to all previously discussed laser-pulsed experiments, and spanning a rather long timescale. Ion irradiation of materials has its own peculiarities, such as an extremely high excitation within extremely narrow nanometric region around an ion trajectory, with the material almost undisturbed between ion tracks [81]. Nonequilibrium electronic transport effects are very important in this case [81]. In our view, it may therefore produce lower values for the coupling parameters similar to the multishot laser experiments discussed above. Thus, for a systematic comparison, coupling estimations within the same experimental methodology with a single ultrashort laser pulse would be beneficial.

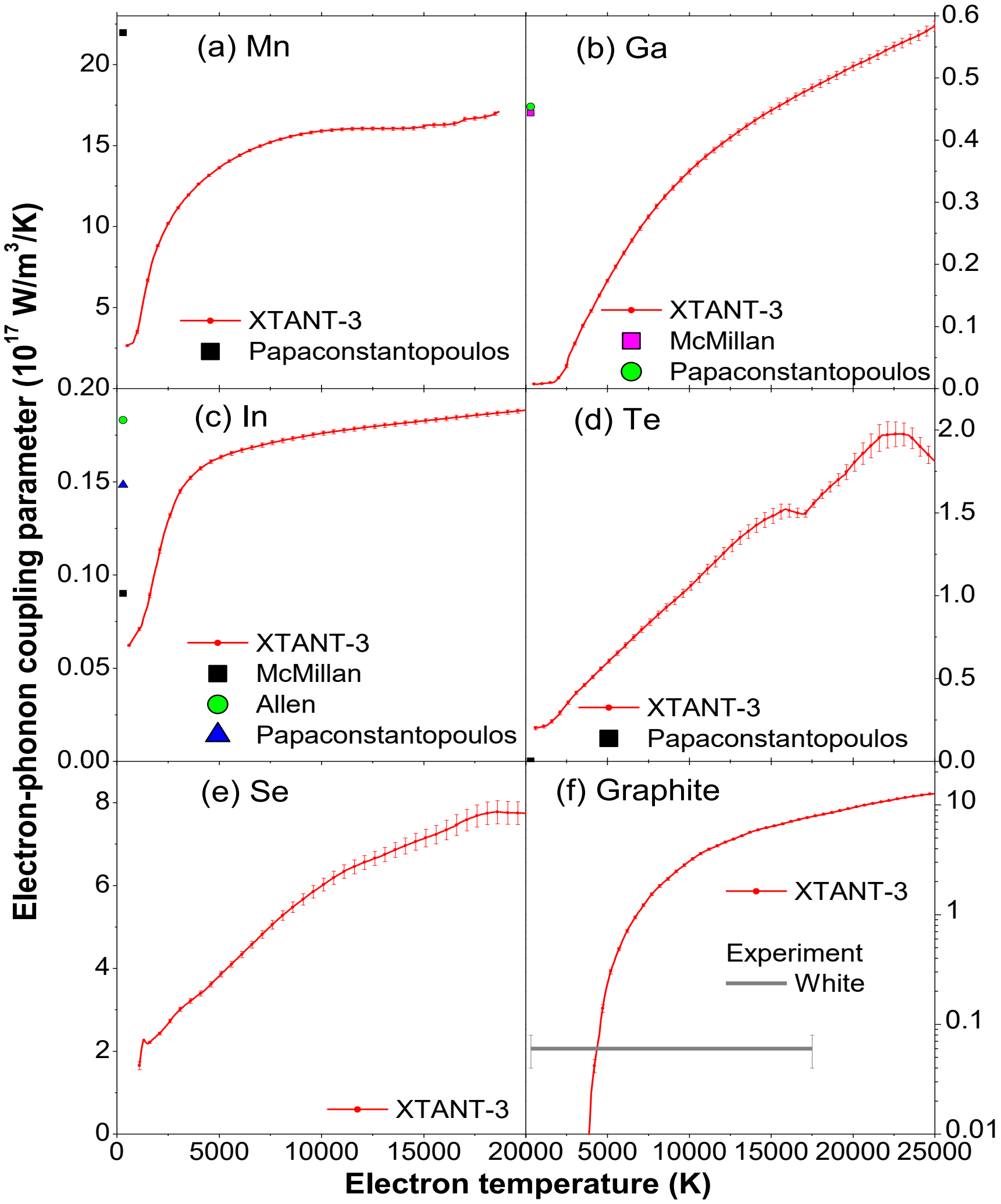

*Figure 11. Electron-phonon coupling parameter in: (a)* ***Manganese****, calculated within XTANT-3 dynamical coupling approach, compared with a low-temperature theoretical estimate by Papaconstantopoulos* [30].
*(b)* ***Gallium****, calculated within XTANT-3 dynamical coupling approach, compared with low-temperature theoretical estimates by McMillan* [29] *and Papaconstantopoulos* [30].
*(c)* ***Indium****, calculated within XTANT-3 dynamical coupling approach, compared with low-temperature theoretical estimates by McMillan* [29]*, Allen et al.* [28]*, and Papaconstantopoulos* [30].
*(d)* ***Tellurium****, calculated within XTANT-3 dynamical coupling approach, compared with a low-temperature theoretical estimate by Papaconstantopoulos* [30].
*(e)* ***Selenium****, calculated within XTANT-3 dynamical coupling approach.*
*(f)* ***Graphite****, calculated within XTANT-3 dynamical coupling approach, compared with experimental data from White et al.* [80].

# IV. Discussion

As we saw in the previous section, experimental data at the room temperature are often scattered and do not agree among themselves, even when obtained with the same experimental method. Different experimental methods almost always produce discrepant results. Many reasons for that can be identified, see e.g. discussion in [40]. One of the important effects experimentally identified in the work [40] was transient nonequilibrium of electrons after a femtosecond laser pulse; an importance of this effect was predicted theoretically e.g. in [82]. Another factor is the temperature of the lattice, which also influences the coupling strength. This effect will be discussed below.

Experiments determining the coupling strength at elevated electronic temperatures are notoriously difficult to perform and interpret [20]. Modern pump-probe experiments measure evolution of the diffraction patterns, which then needs to be translated into the information on the atomic temperature, and then connected to the electron-ion coupling factor. Such interpretations are usually model-dependent, as the two-temperature model (TTM) needs to be used to determine some of the parameters that are not directly accessible in the experiments (see e.g. supplementary information in [20]). There are typically many approximations introduced along the way of extracting the coupling strength from the experimental diffraction patterns, such as, on the spatial profile of the energy deposition by the pump pulse; an approximation of instant thermalization of electrons; unchanged atomic density and pressure during the experiments; unchanged interatomic potential upon electronic excitation; an instant thermalization of atoms. As was shown in multiple works [11,19,40,82], any and all of these approximations may not be valid under intense ultrashort irradiations. It makes it difficult to compare a calculated electron-ion (electron-phonon) coupling strength to the experimental one. Future dedicated theoretical works should focus on providing the results that can be directly compared to the experimental ones: e.g. diffraction patterns evolution.

In most of the cases presented above in the Results section, calculations based on the Eliashberg spectral function formalism, or similar approaches based on the calculations of the matrix elements, can be matched reasonably well to our results by rescaling with a constant coefficient (typically a factor of 1/4). There may be various reasons for the need of such a rescaling, as discussed e.g. in the original work on example of aluminum [16]. Petrov *et al*. in [17] noticed that reducing the effective electron mass reduces the coupling strength, as they demonstrated for gold. It was also suggested by Gorbunov *et al*. that a stronger electron screening would reduce the coupling parameter [24]; a similar notion was discussed by Petrov *et al*. [17]. The recent work by Waldecker *et al*. [19] suggested that the electron-phonon coupling can be reduced due to the fact that different phonon modes couple differently to electrons – the fact that is naturally accounted for in our XTANT-3 calculations. Accounting for the different coupling to different phonon modes allowed the authors of [19] to achieve an agreement with the experimental data in aluminum, the result which was not possible to achieve in other *ab-initio* calculations without rescaling.

We note that a qualitative agreement of the dependencies of the electron-phonon coupling parameter on the electronic temperature among various approaches in most of the studied materials suggests that any of them can be used in practice, provided a benchmark against experimental data is available to allow for a proper rescaling if required. The exceptions appear to be magnetic materials: Ni, Fe, Pt, where the behavior of the coupling parameter is qualitatively different in our approach compared to other calculations, which poses an interesting question for future studies.

In certain cases, such as Pb, V, Nb, Ru, W, our results do not agree well with the experimental data at room temperatures. We deliberately avoided rescaling or any other manipulation of our results even when they disagree with the experimental data. Such cases of discrepancies demonstrate that there are still open questions to clarify, either theoretically or experimentally (or both). As one of the possibilities, it would be important to check other tight binding parameterizations. Unfortunately, there is currently no available transferrable TB parameterization for those solids, to the best of our knowledge. We hope that the results reported here will motivate the readers for future dedicated work in both directions: performing new experiments with ever-increasing precision and covering high-electron-temperature range, and further developing the theory and identifying any important effects missing.

## IV.1. Trends in electron-ion coupling across the periodic table

The obtained data allow us to comparatively analyze the electron-ion coupling in elemental solids across the periodic table. We identified the overall trends, shown in Figure 12. This figure demonstrates the coupling parameter calculated with XTANT-3, normalized per number of electrons in the conduction band of the material, at $T_e$ = 10000 K representing an intermediate value within the studied intervals of the electronic temperatures. We can identify the overall trend of the coupling strength to decrease with increase of the atomic number. This effect is attributed to the fact that with the increase of the atomic mass, inertia increases, reducing the atomic velocity at the same kinetic energy. As the coupling parameter depends on the atomic velocity (see next section), it leads to the decrease of the coupling strength with the increasing atomic number. In other words, with the increasing atomic mass, the Born-Oppenheimer approximation holds better and better.

Next, we divided elements in groups by the bonding types, depending on to which shell the majority of the conduction band electrons belongs (*s*, *p* or *d*). This allowed us to identify additional trends in the coupling parameter in the materials where the outermost *d*-shells are being filled: the 3d, 4d and 5d groups. Within each group, highlighted in Figure 12 with dashed lines, we see an initial increase of the coupling parameter (taken at $T_e$ = 10000 K), peaking at the middle of filling a *d*-shell and decreasing after that. The data points resemble Gaussian peaks, shown in the figure to guide the eye. For example, within the 5$^{th}$ period elements (4d electron shell) the trend starts with Sr (limiting case of no electrons in the 4d shell), has a peak for Tc (5 electrons in the 4d shell) and finishes with Cd (10 electrons in the 4d shell, i.e. fully occupied). Similar trends are observed for the 4$^{th}$ and 6$^{th}$ period elements, 3d and 5d shells, respectively. The case of iron deviates from the trend, exhibiting a large coupling parameter (in fact, the largest among all the elements considered in this work).

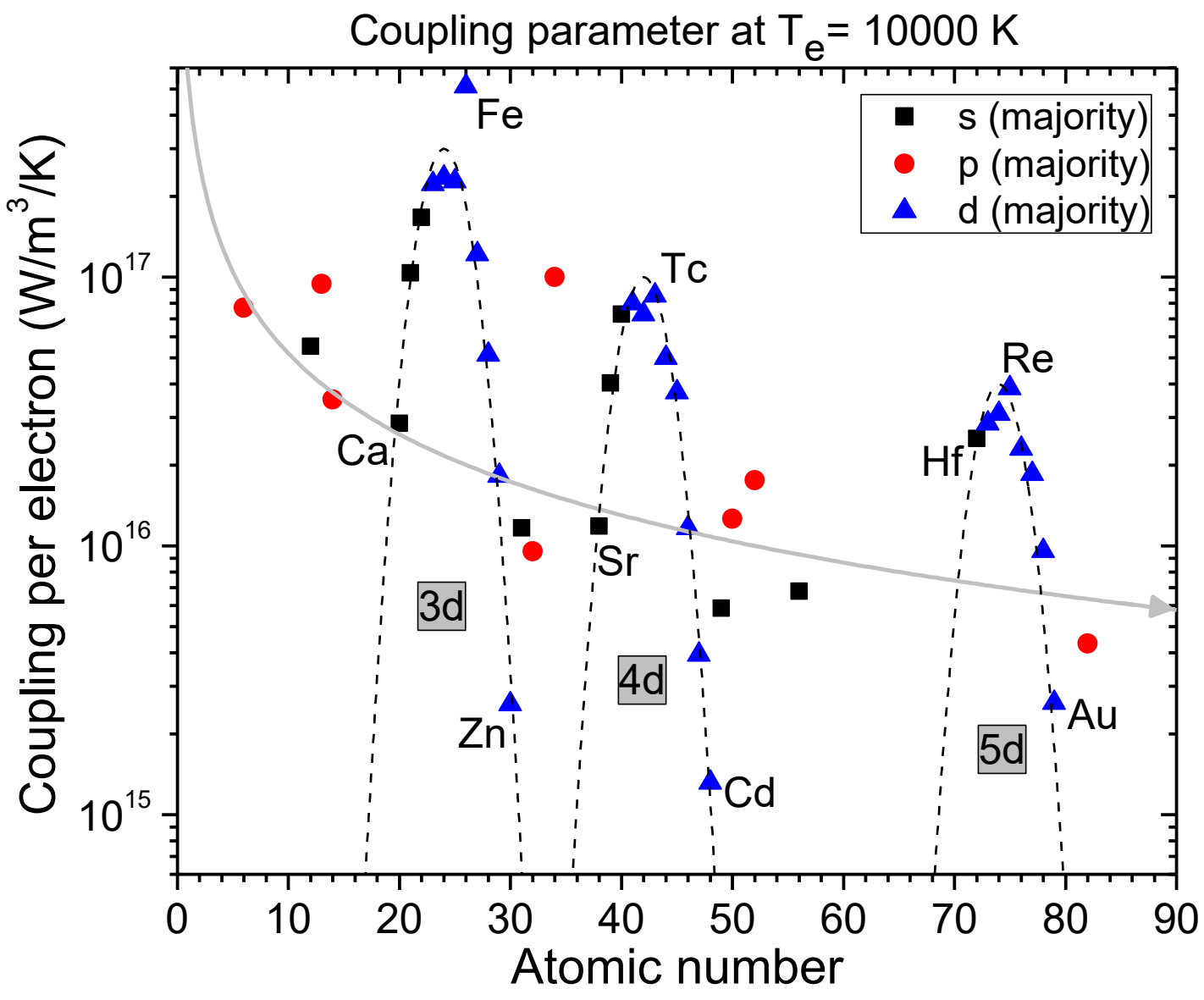

*Figure 12. Electron-phonon coupling parameter normalized to the number of electrons in the conduction band of the corresponding material at electronic temperature of $T_e$=10000 K as a function of the atomic number (Z) in the periodic table, calculated within XTANT-3 dynamical coupling approach. Grey arrow indicates inverse dependence ~1/Z. Dashed lines are Gaussian peaks centered at Z=24, 42, 74. The elements are divided in groups based on the bonding type, i.e. depending on to which shell the majority of the conduction band electrons belongs (s, p or d).*

The coupling parameter is less sensitive to the crystal structure than to the electronic population and, correspondingly, to the atomic number. It is shown on the example of two different solid phases of chromium: Im$\bar{3}$m and Pm$\bar{3}$n phase, Figure 13. The difference in coupling between the two phases is about 15%, except for the low electronic temperatures where the curves diverge more significantly. This result also indicates that experimental measurements of the coupling strength at near-room temperatures may be rather sensitive to the particular atomic structure of the material, with possible effects of its purity, crystallinity and crystallite sizes.

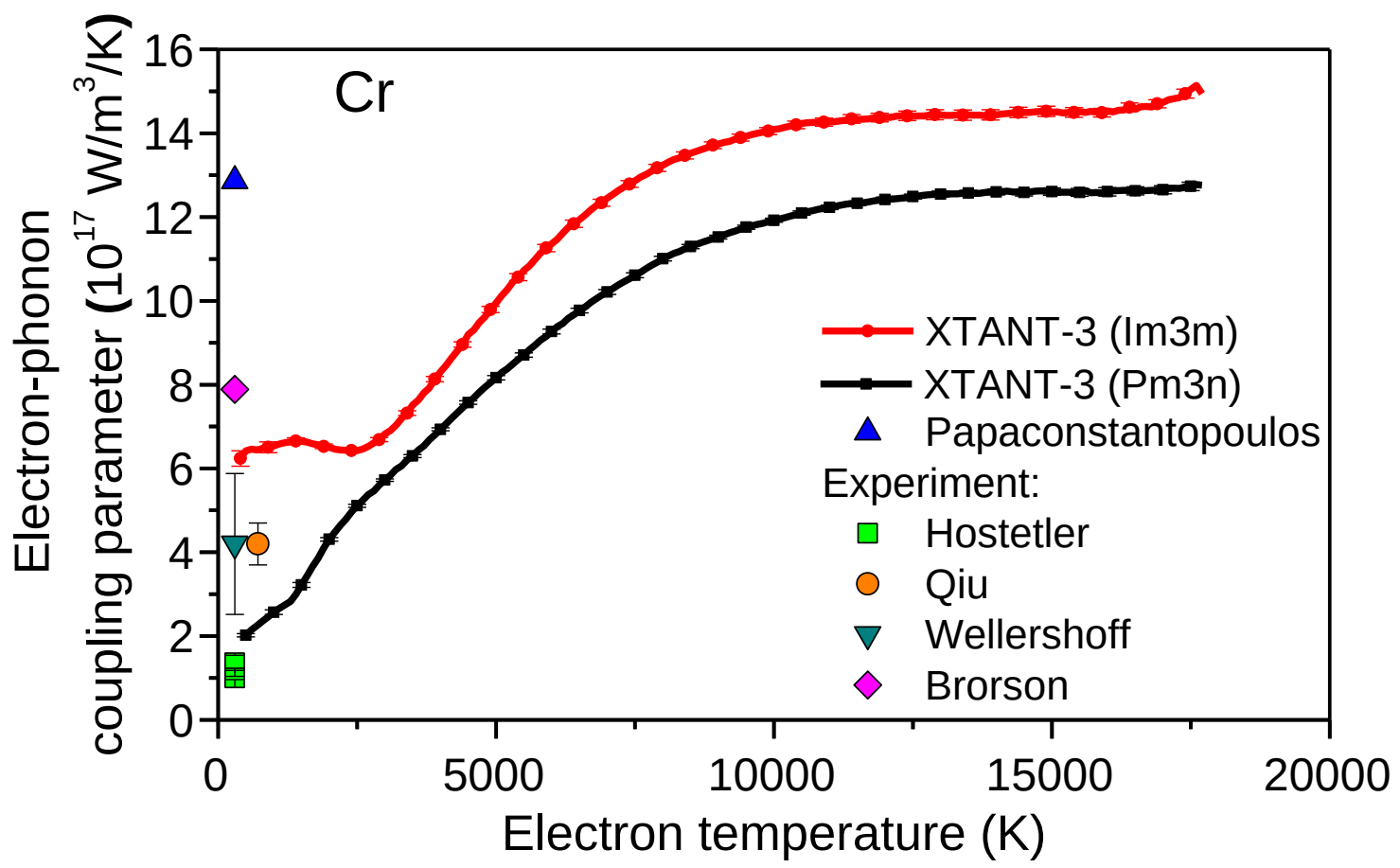

*Figure 13. Electron-phonon coupling parameter in chromium in Im$\bar{3}$m phase (reproduced from Figure 8 above), and in Pm$\bar{3}$n phase, calculated within XTANT-3 dynamical coupling approach.*

## IV.2. Electron-phonon coupling dependence on atomic temperature

As was discussed e.g. in [23], and in the original work on the non-adiabatic coupling by Tully *et al*. [83,84], the electron-ion matrix element within the first order approximation can be written as follows: $\langle \psi_j(t) | \psi_i(t_0) \rangle \sim \delta t \dot{\vec{R}} \vec{d}$, where $\dot{\vec{R}}$ is the atomic velocity, and $\vec{d}$ is the non-adiabatic coupling vector. Thus, from Eq.(2) it follows that the coupling factor scales linearly with the atomic temperature:

$$G(T_e, T_a) \sim \dot{\vec{R}}^2 \sim T_a \tag{8}$$

Such an estimation of the linear dependence of the coupling on the atomic temperature is, of course, very crude and does not account for the dependence of the non-adiabatic coupling vector on the atomic temperature, structure, pressure, density etc., but it can be used as a rule of thumb for $G(T_e, T_a)$, at least below the melting temperature of the material. A nearly linear dependence was already reported for silicon [21].

Figure 14 shows the coupling strength of gold, chromium in Im$\bar{3}$m phase and in Pm$\bar{3}$n phase calculated with XTANT-3 at various atomic temperatures. Rescaling the curves by a constant factor (as shown in the bottom panels of Figure 14) brings them very close together, with only small deviations at the low electronic temperatures. It is to be expected, since the coupling parameter is more sensitive to the electronic band structure at low electronic temperatures, and thus stronger depends on the atomic temperature. The rescaling factor as a function of the atomic temperature is shown in Figure 15, where we can see the points following an almost linear dependence, for all the three cases reported here (and the same result was obtained for silicon in [21]). It confirms that the coupling parameters dependence on the atomic temperature can be approximated as linear. It seems to be a universal behavior, independent of the material and its phase and structure. There is an indication of a saturation at high atomic temperatures, where the scaling factor deviates from the linear dependence, showing that above the melting temperature ($T_{melt}$ = 1337 K for gold at normal conditions, for example) the approximation (8) does not hold so well.

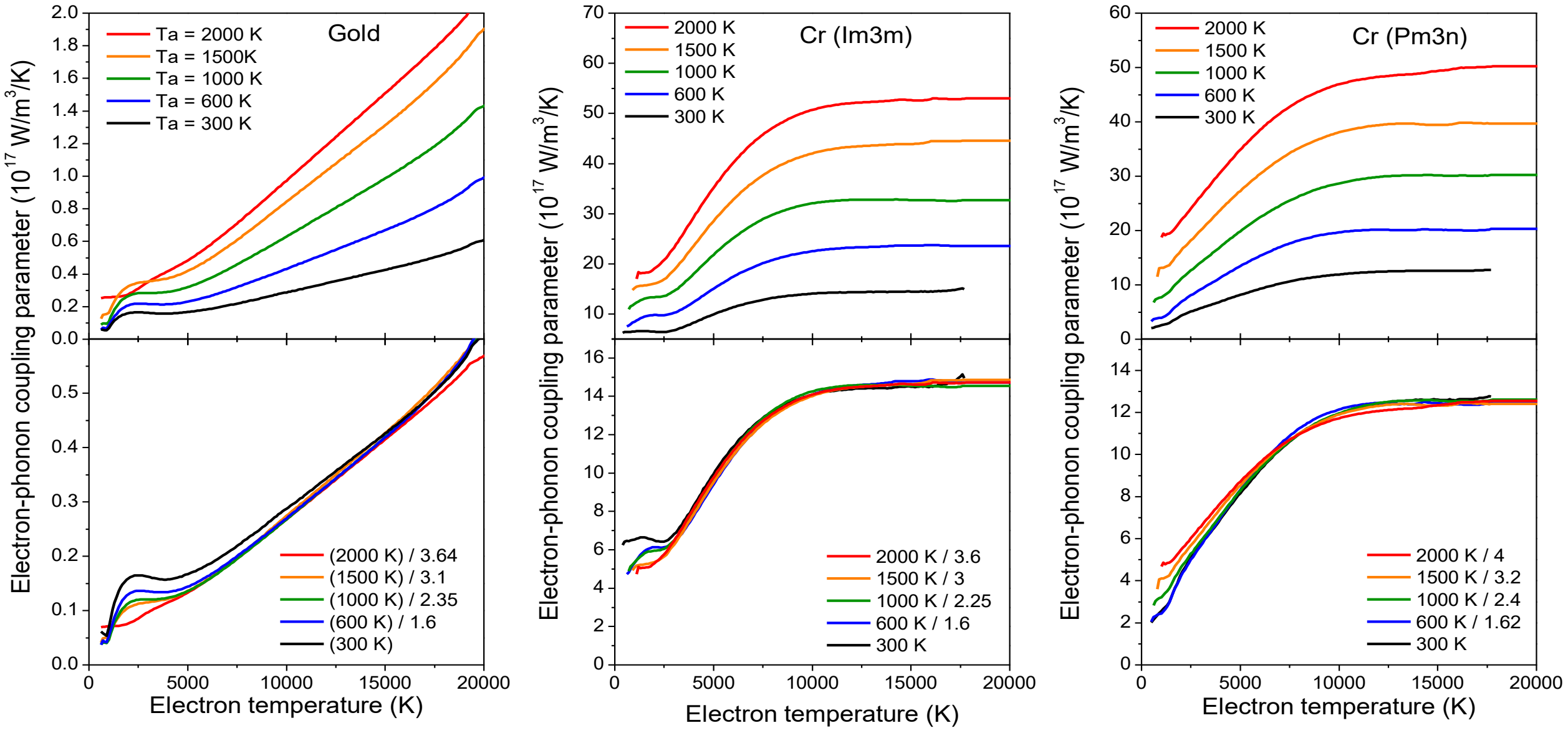

*Figure 14. Electron-phonon coupling parameter in gold (left panel), chromium in Im$\bar{3}$m phase (middle panel), and chromium in Pm$\bar{3}$n phase (right panel) calculated within XTANT-3 dynamical coupling approach for different atomic temperatures (top panel), and rescaled by a factor listed in the legend to match the room temperature coupling (bottom panel).*

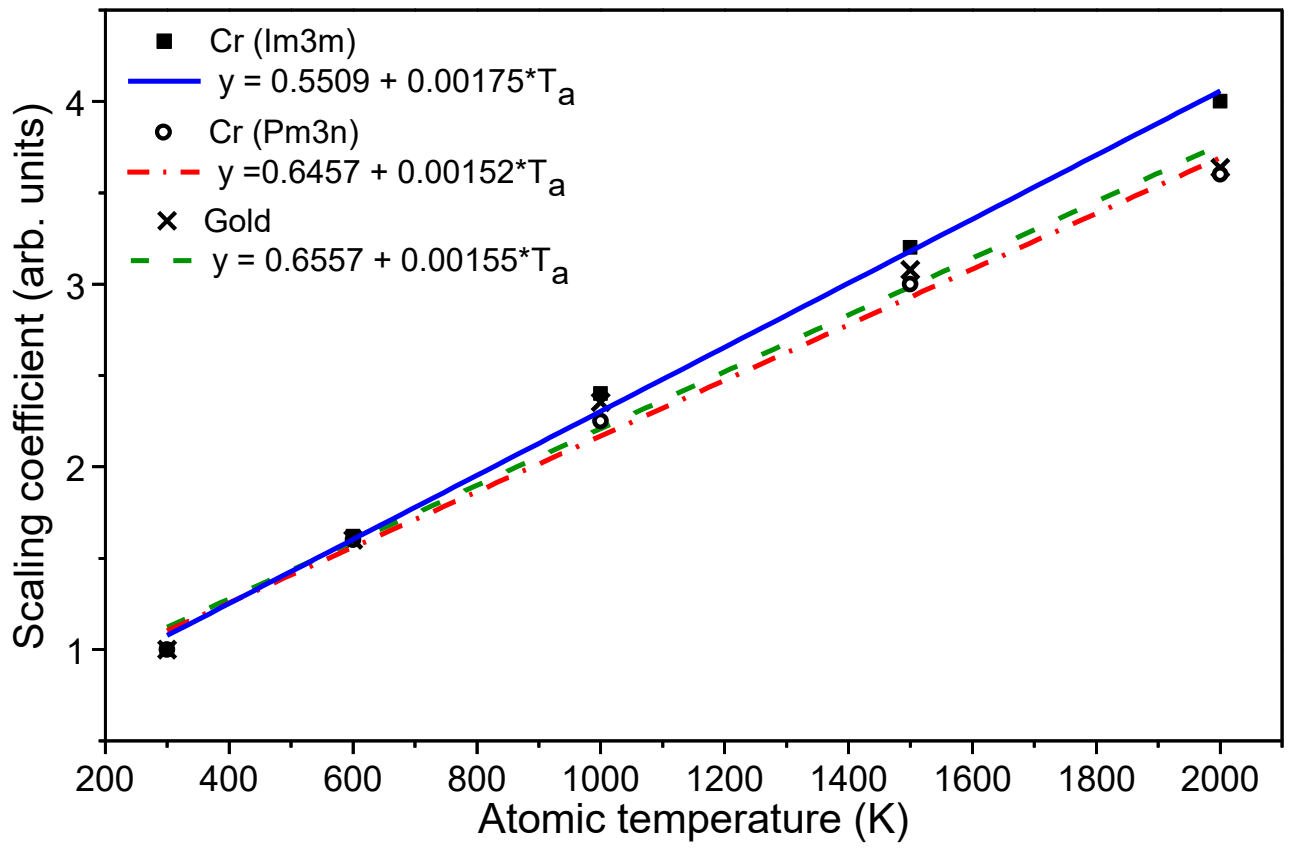

*Figure 15. Scaling parameter for the electron-phonon coupling parameter in gold, chromium in Im$\bar{3}$m phase, and chromium in Pm$\bar{3}$n phase at different atomic temperatures that rescales the curves to coincide with that at $T_a$=300K (see Figure 14). The lines are linear fits through the data points.*

## IV.3. Electron-phonon coupling dependence on density

We analyzed the dependence of the electron-phonon coupling strength on the material density by rescaling the volume of the simulation cell keeping the same atomic configuration on an example of gold. Figure 16 shows the coupling parameter in gold calculated for different supercell volumes from 0.8 to 1.2 of the normal volume $V_0$. The coupling parameter scales almost linearly up to ~1.1 $V_0$, as can be seen in Figure 16. Rescaling by a constant factor makes the curves lie close to each other, with some divergence starting at high electronic temperatures for an expanded supercell corresponding to a reduced material density. It shows a linear reduction of coupling with expansion of material (see Figure 17), or, in other words, a linear scaling of the coupling parameter with the material density in a solid state.

At volumes exceeding ~1.1 $V_0$, lattice instabilities occur, especially at elevated electronic temperatures, as was studied in detail in our previous work (to be published elsewhere). This, in turn, leads to a change of the slope of the coupling curve seen in Figure 16.

As one can see, electron-phonon (electron-ion) coupling strength is a function of many variables, the fact that has to be taken into account when designing new experiments aimed to measure the coupling strength, and also when analyzing the experimental results within a framework of any theoretical model.

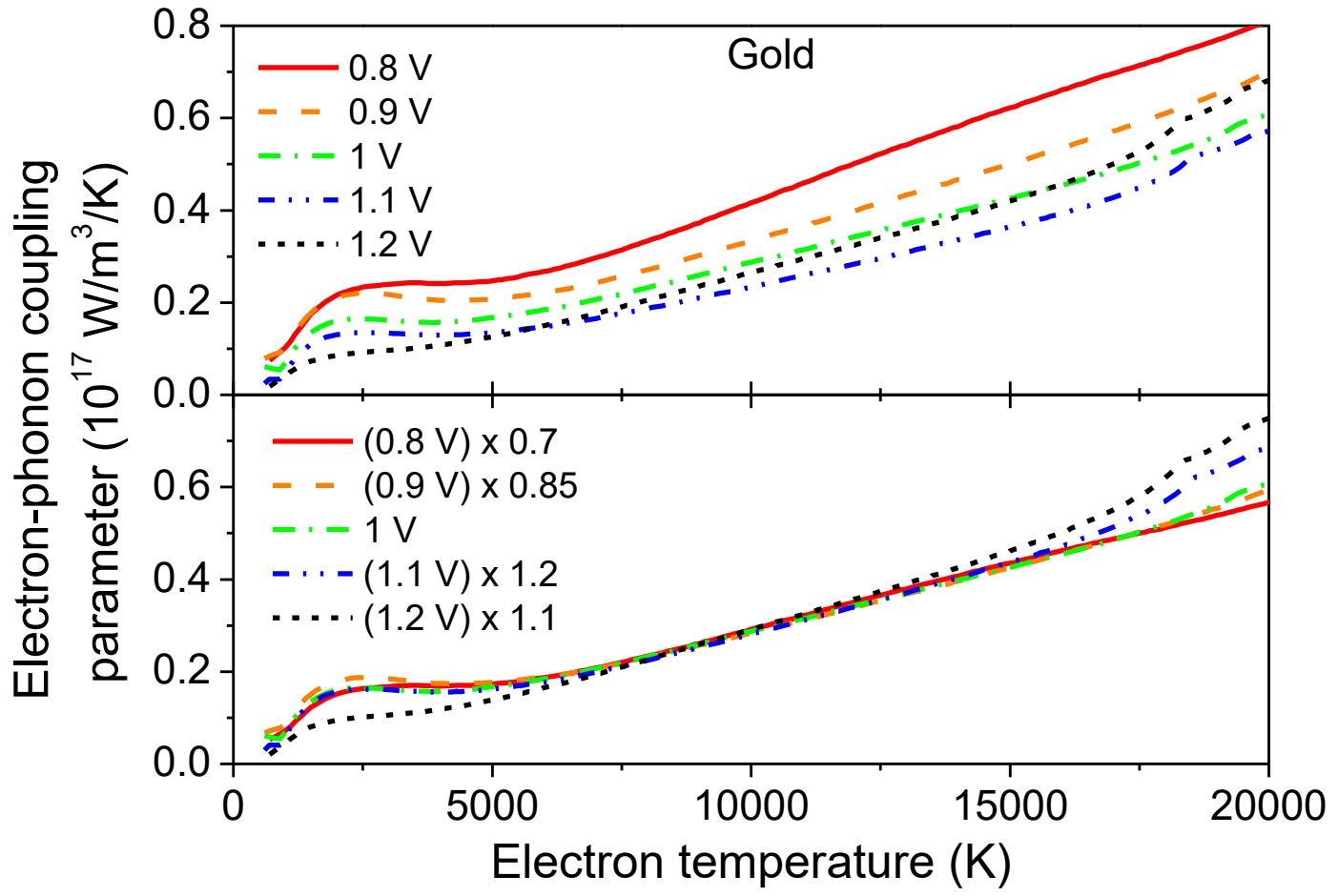

*Figure 16. Electron-phonon coupling parameter in gold calculated within XTANT-3 dynamical coupling approach for different volumes of the supercell (top panel), and rescaled by a factor listed in the legend to match the normal volume coupling (bottom panel).*

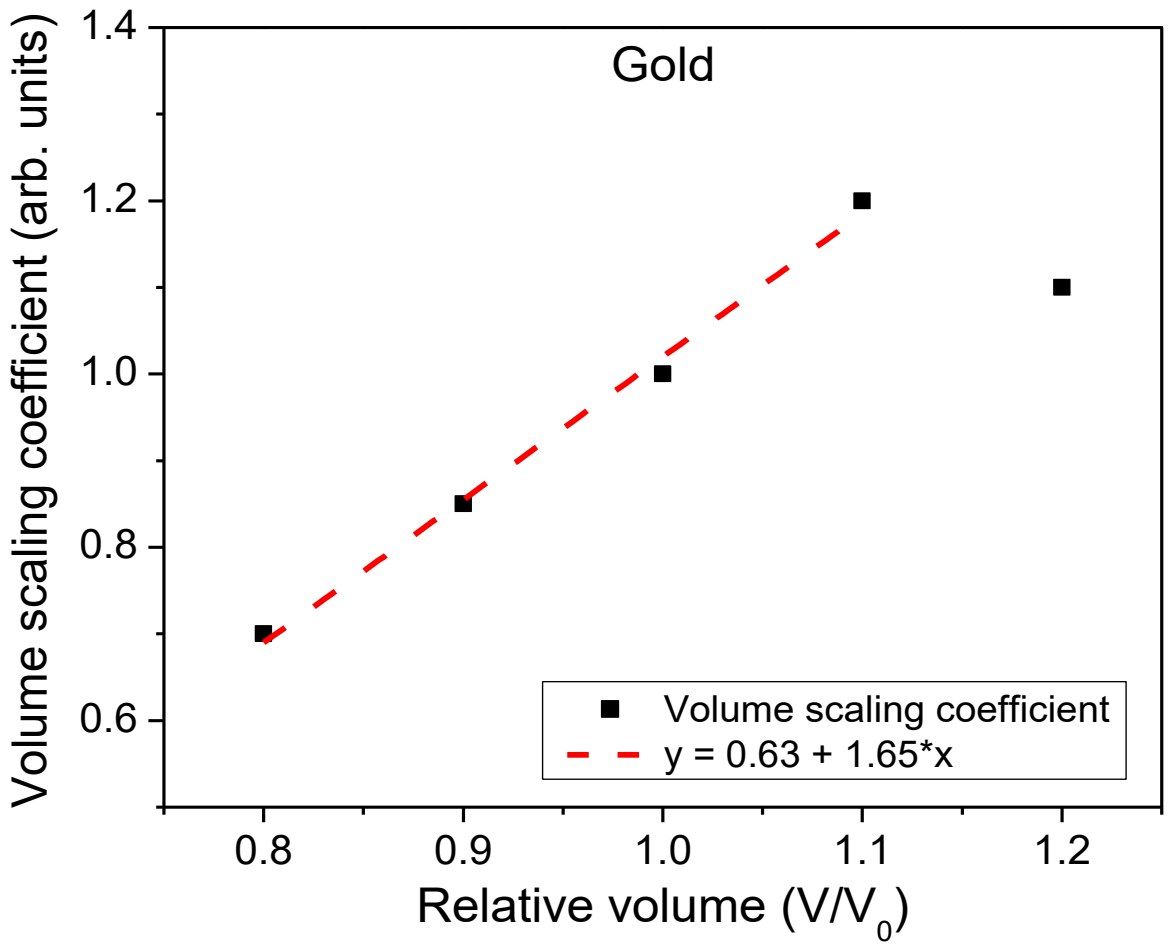

*Figure 17. Scaling parameter for the electron-phonon coupling parameter in gold for different supercell volumes that rescales the curves to coincide with that at normal volume (see Figure 14). The line is a linear fit through the data points (except for the one at 1.2$V_0$).*

# V. Conclusions

A nonperturbative scheme for calculations of the nonadiabatic electron-ion (electron-phonon) coupling was presented. It is especially well suited for the implementations into the tight-binding molecular dynamics, as illustrated with the application of the XTANT-3 code. The model is, in principle, applicable to any atomic configuration and motion: it is not limited to harmonic potential and periodic crystals (phononic approximation).

With this model, the electron-phonon coupling strengths for many elemental materials were calculated. The results compare well with the available experimental data at high-electronic temperatures. In many cases, the agreement is reasonable also with near-room-temperature data, with some exceptions: in Pb, V, Nb, Ru, and W, our calculations predicted a significantly lower coupling strength than the experimental one at low electronic temperatures. Those cases require future dedicated studies.

The presented results often qualitatively agree with the calculations based on the Eliashberg function formalism, suggesting that either of the methods can be used with a proper rescaling, when required. In a few cases, such as Ni, Fe, and Pt, XTANT-3 results deviate from the previously published calculations qualitatively. It indicates that some important effects are possibly missing in the Eliashberg formalism, such as the dependence of the interatomic potential and other atomic properties including electron-ion scattering matrix elements, atomic vibrational spectra, and electronic band structure on both electronic and atomic temperature. It includes such effects as phonon hardening and displacive excitation of coherent phonons. In contrast, the presented model naturally accounts for all the nonthermal effects and their synergetic interplay with the electron-ion coupling.

A general trend of the electron-phonon coupling decrease with the increasing atomic number (and hence atomic mass) was identified. A more detailed analysis showed additional trends with respect to the level of outermost d-shell occupation for the groups of elements from the $4^{th}$, $5^{th}$ and $6^{th}$ periods in the periodic table. The coupling parameter increases with the increasing number of outermost d-shell electrons up to a half-filled d-shell, and then decreases until a shell is completely filled.

Linear dependencies of the electron-phonon coupling on the atomic temperature and on the material density were demonstrated. This fact implies that experimental measurements of the coupling parameter must be performed with a careful control of other parameters, and reported at least as a function of both, electronic as well as atomic temperatures. As material density also plays a role, it must be one of the controlled parameters in the experiments.

The performed analysis demonstrates a clear lack of experimental data of the materials electron-phonon coupling strength measured at high electron temperatures. Without such experiments, it is hardly possible to make a solid conclusion about the accuracy of various theoretical models.

## Acknowledgements

The authors gratefully acknowledge financial support from the Czech Ministry of Education, Youth and Sports (Grants No. LTT17015, No. EF16_013/0001552, and No. LM2015083). I. Milov gratefully acknowledges support from the Industrial Focus Group XUV Optics of the MESA+ Institute for Nanotechnology of the University of Twente; the industrial partners ASML, Carl Zeiss SMT GmbH, and Malvern Panalytical, the Province of Overijssel, and the Netherlands Organisation for Scientific Research (NWO).

## Appendix. Convergence study

We ensured the convergence of our results with respect to the MD timestep, which was achieved for timesteps below ~2 fs at the near-room temperature modeling, see Figure 18. Further decrease of the timestep below 2 fs results only in a small difference. Increasing the atomic temperature generally requires reducing the timestep of simulation.

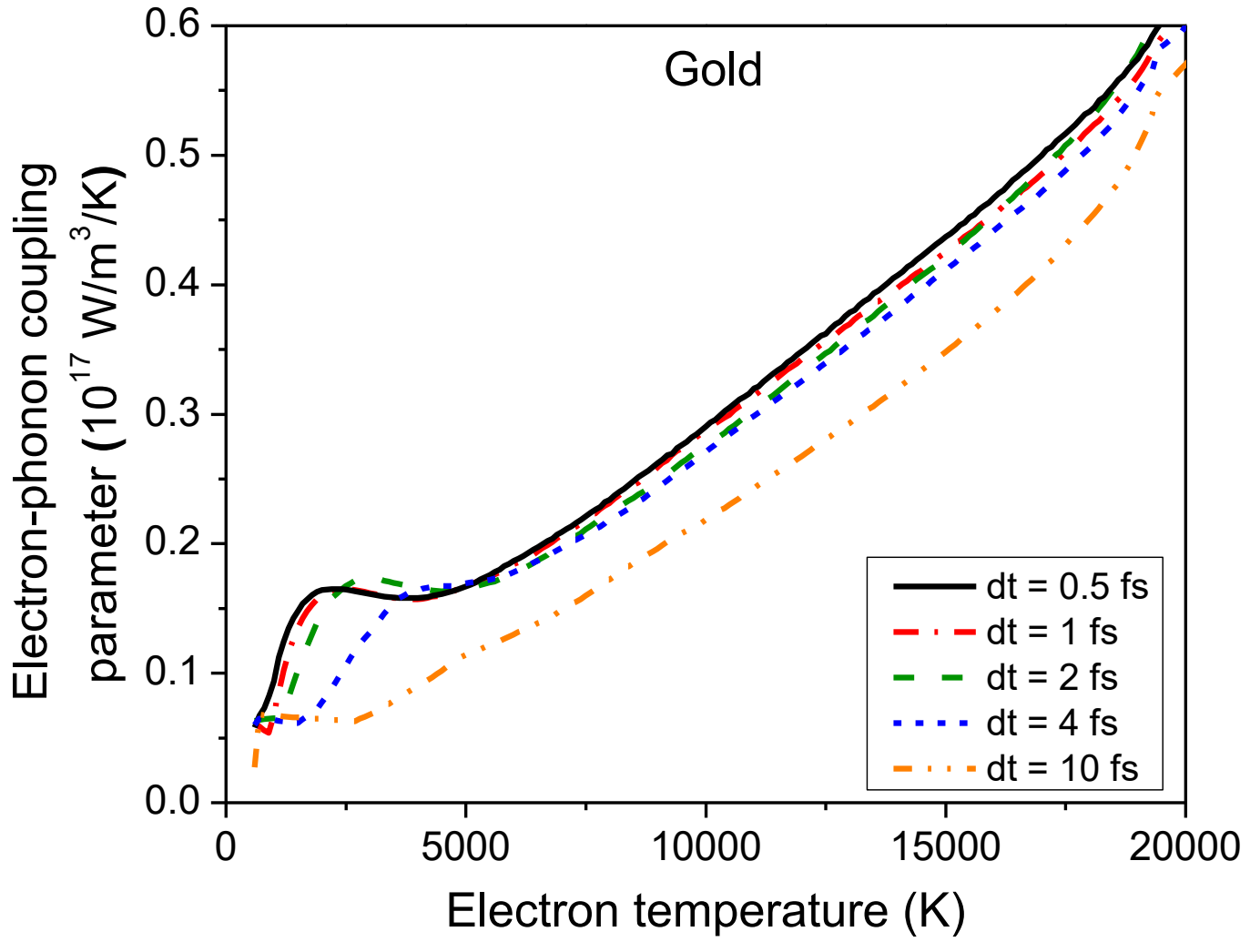

*Figure 18. Electron-phonon coupling parameter in gold calculated within XTANT-3 dynamical coupling approach at the near-room temperature with different timesteps.*

The supercell size must be above ~200 atoms for the tight-binding molecular dynamics simulations with a single gamma-point dynamics, as reported in [21]. It is shown in Figure 19 on an example of two materials and different numbers of atoms in a supercell. The results are converged for about 200-300 atoms, and the convergence is faster for higher electronic temperatures. Please note that these calculations are performed for a single gamma point in the Brillouin zone, as written in the main text.

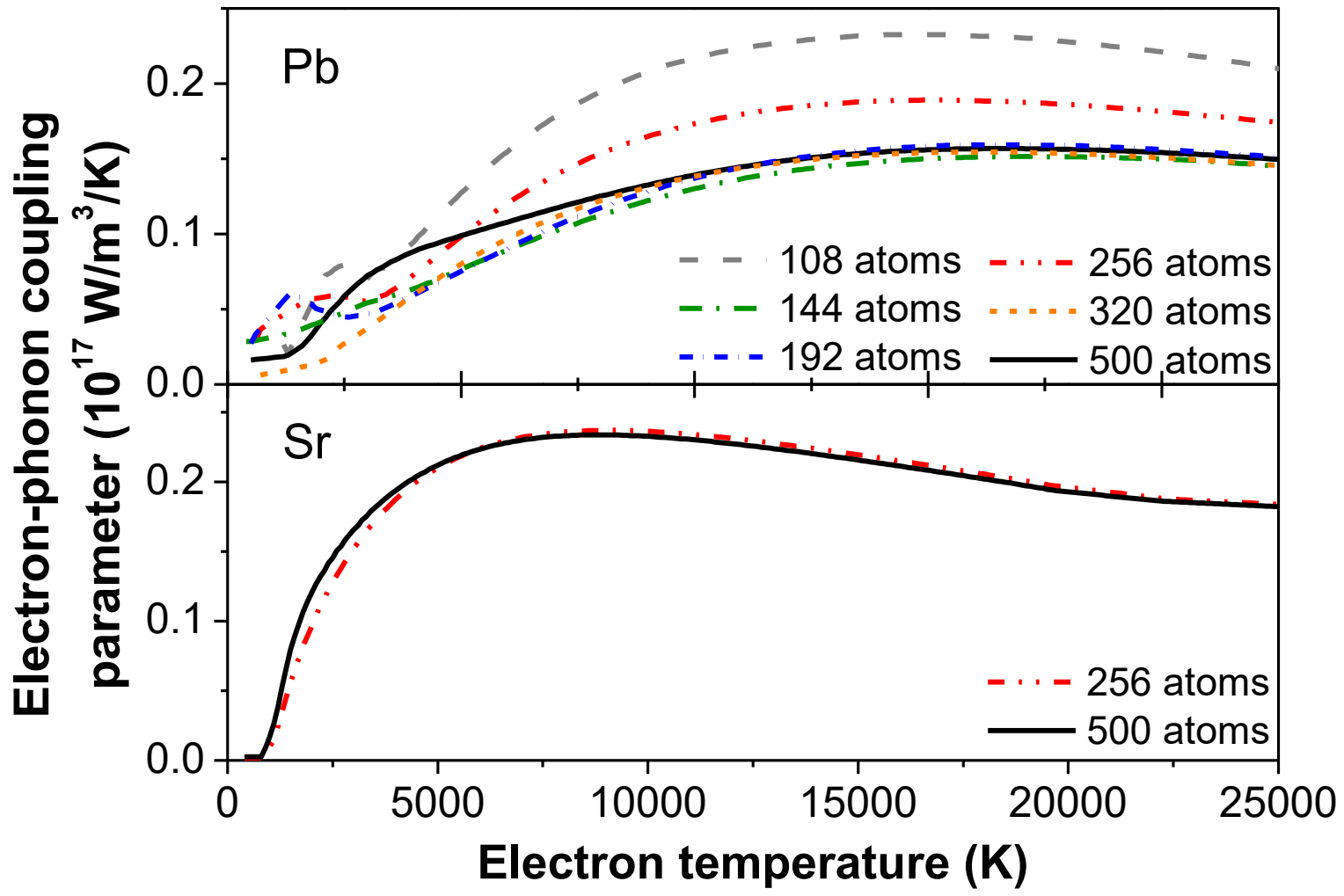

*Figure 19. Electron-phonon coupling parameter in lead (upper panel) and strontium (lower panel) calculated within XTANT-3 dynamical coupling approach for different numbers of atoms in the supercell.*

It is, in principle, possible to calculate the electron-phonon transition probabilities for multiple *k*-points, which is a rigorous and straightforward way to improve the precision. It is expected to improve a convergence at low electronic temperatures, and at smaller numbers of atoms in a simulation box, but it will increase the calculation time. However, this improvement of precision is beyond the scope of the present work. It can be considered as a next step of research, after a larger number of reliable and accurate experimental data will become available.

# References

[1] W. E. Lamb, W. P. Schleich, M. O. Scully, and C. H. Townes, Rev. Mod. Phys. **71**, S263 (1999).

[2] S. Y. Kruchinin, F. Krausz, and V. S. Yakovlev, Rev. Mod. Phys. **90**, 021002 (2018).

[3] G. Amoako, Appl. Phys. Res. **11**, 1 (2019).

[4] K. Sugioka and Y. Cheng, Light Sci. Appl. **3**, e149 (2014).

[5] H.-P. Berlien, G. J. Müller, H. Breuer, N. Krasner, T. Okunata, and D. Sliney, editors , *Applied Laser Medicine* (Springer-Verlag, Berlin Heidelberg, 2003).

[6] M. Braun, P. Gilch, and W. Zinth, editors , *Ultrashort Laser Pulses in Biology and Medicine* (Springer-Verlag, Berlin Heidelberg, 2008).

[7] R. E. Slusher, Rev. Mod. Phys. **71**, S471 (1999).

[8] M. V. Shugaev, C. Wu, O. Armbruster, A. Naghilou, N. Brouwer, D. S. Ivanov, T. J. Y. Derrien, N. M. Bulgakova, W. Kautek, B. Rethfeld, and L. V. Zhigilei, MRS Bull. **41**, 960 (2016).

[9] B. Rethfeld, D. S. Ivanov, M. E. Garcia, and S. I. Anisimov, J. Phys. D. Appl. Phys. **50**, 193001 (2017).

[10] M. V. Shugaev, M. He, S. A. Lizunov, Y. Levy, T. J.-Y. Derrien, V. P. Zhukov, N. M. Bulgakova, and L. V. Zhigilei, in *Adv. Appl. Lasers Mater. Sci. Springer Ser. Mater. Sci. Vol 274.*, edited by P. Ossi (Springer, Cham, 2018), pp. 107–148.

[11] N. Medvedev, V. Tkachenko, V. Lipp, Z. Li, and B. Ziaja, 4open **1**, 3 (2018).

[12] V. L. Shabansky and V. P. Ginzburg, Reports Acad. Sci. USSR **100**, 445 (1955).

[13] M. I. Kaganov, I. M. Lifshitz, and L. V. Tanatarov, Sov. Phys. JETP **4**, 173 (1957).

[14] S. I. Anisimov, B. L. Kapeliovich, and T. L. Perel-man, J. Exp. Theor. Phys. **39**, 375 (1974).

[15] D. Ivanov and L. Zhigilei, Phys. Rev. B **68**, 064114 (2003).

[16] Z. Lin, L. Zhigilei, and V. Celli, Phys. Rev. B **77**, 075133 (2008).

[17] Y. V. Petrov, N. A. Inogamov, and K. P. Migdal, JETP Lett. **97**, 20 (2013).

[18] A. M. Brown, R. Sundararaman, P. Narang, W. A. Goddard, and H. A. Atwater, Phys. Rev. B **94**, (2016).

[19] L. Waldecker, R. Bertoni, R. Ernstorfer, and J. Vorberger, Phys. Rev. X **6**, 021003 (2016).

[20] M. Z. Mo, Z. Chen, R. K. Li, M. Dunning, B. B. L. Witte, J. K. Baldwin, L. B. Fletcher, J. B. Kim, A. Ng, R. Redmer, A. H. Reid, P. Shekhar, X. Z. Shen, M. Shen, K. Sokolowski-Tinten, Y. Y. Tsui, Y. Q. Wang, Q. Zheng, X. J. Wang, and S. H. Glenzer, Science **360**, 1451 (2018).

[21] N. Medvedev, Z. Li, V. Tkachenko, and B. Ziaja, Phys. Rev. B **95**, 014309 (2017).

[22] J. Daligault and D. Mozyrsky, Phys. Rev. B **98**, (2018).

[23] N. Medvedev, Z. Li, and B. Ziaja, Phys. Rev. B **91**, 054113 (2015).

[24] S. A. Gorbunov, N. A. Medvedev, P. N. Terekhin, and A. E. Volkov, Nucl. Instruments Methods Phys. Res. Sect. B Beam Interact. with Mater. Atoms **354**, 220 (2015).

[25] D. A. Papaconstantopoulos and M. J. Mehl, J. Phys. Condens. Matter **15**, R413 (2003).

[26] N. Medvedev, in *Opt. Damage Mater. Process. by EUV/X-Ray Radiat. VII*, edited by L. Juha, S. Bajt, and S. Guizard (SPIE, 2019), p. 25.

[27] N. Medvedev, Appl. Phys. B **118**, 417 (2015).

[28] P. B. Allen and M. L. Cohen, Phys. Rev. **187**, 525 (1969).

[29] W. L. McMillan, Phys. Rev. **167**, 331 (1968).

[30] D. A. Papaconstantopoulos, *Handbook of the Band Structure of Elemental Solids* (Springer US, Boston, MA, 2015).

[31] B. A. Sanborn, P. B. Allen, and D. A. Papaconstantopoulos, Phys. Rev. B **40**, 6037 (1989).

[32] B. Hüttner and G. Rohr, Appl. Surf. Sci. **103**, 269 (1996).

[33] J. L. Hostetler, A. N. Smith, D. M. Czajkowsky, and P. M. Norris, Appl. Opt. **38**, 3614 (1999).

[34] Z. Li-Dan, S. Fang-Yuan, Z. Jie, and T. Da-Wei, Acta Phys. Sin **61**, 134402 (2012).

[35] B. Y. Mueller and B. Rethfeld, Phys. Rev. B **87**, 035139 (2013).
[36] J. Hohlfeld, J. G. Mueller, S.-S. Wellershoff, and E. Matthias, Appl. Phys. B Lasers Opt. **64**, 387 (1997).
[37] A. Nakamura, T. Shimojima, M. Nakano, Y. Iwasa, and K. Ishizaka, Struct. Dyn. **3**, (2016).
[38] S. D. Brorson, A. Kazeroonian, J. S. Moodera, D. W. Face, T. K. Cheng, E. P. Ippen, M. S. Dresselhaus, and G. Dresselhaus, Phys. Rev. Lett. **64**, 2172 (1990).
[39] Q. Zheng, X. Shen, K. Sokolowski-Tinten, R. K. Li, Z. Chen, M. Z. Mo, Z. L. Wang, S. P. Weathersby, J. Yang, M. W. Chen, and X. J. Wang, J. Phys. Chem. C **122**, 16368 (2018).
[40] A. Giri, J. T. Gaskins, B. M. Foley, R. Cheaito, and P. E. Hopkins, J. Appl. Phys. **117**, (2015).
[41] T. G. White, P. Mabey, D. O. Gericke, N. J. Hartley, H. W. Doyle, D. McGonegle, D. S. Rackstraw, A. Higginbotham, and G. Gregori, Phys. Rev. B **90**, 014305 (2014).
[42] L. Guo and X. Xu, J. Heat Transfer **136**, (2014).
[43] N. A. Smirnov, Phys. Rev. B **101**, 094103 (2020).
[44] K. P. Migdal, Y. V. Petrov, and N. A. Inogamov, in *Fundam. Laser-Assisted Micro-Nanotechnologies 2013*, edited by V. P. Veiko and T. A. Vartanyan (SPIE, 2013), p. 906503.
[45] X. Y. Wang, D. M. Riffe, Y.-S. Lee, and M. C. Downer, Phys. Rev. B **50**, 8016 (1994).
[46] V. Recoules, J. Clérouin, G. Zérah, P. M. Anglade, and S. Mazevet, Phys. Rev. Lett. **96**, 055503 (2006).
[47] T. Q. Qiu and C. L. Tien, Int. J. Heat Mass Transf. **35**, 719 (1992).
[48] P. B. Corkum, F. Brunel, N. K. Sherman, and T. Srinivasan-Rao, Phys. Rev. Lett. **61**, 2886 (1988).
[49] H. E. Elsayed-Ali, T. B. Norris, M. A. Pessot, and G. A. Mourou, Phys. Rev. Lett. **58**, 1212 (1987).
[50] M. Z. Mo, V. Becker, B. K. Ofori-Okai, X. Shen, Z. Chen, B. Witte, R. Redmer, R. K. Li, M. Dunning, S. P. Weathersby, X. J. Wang, and S. H. Glenzer, Rev. Sci. Instrum. **89**, 10 (2018).
[51] G. M. Choi, C. H. Moon, B. C. Min, K. J. Lee, and D. G. Cahill, Nat. Phys. **11**, 576 (2015).
[52] B. I. Cho, T. Ogitsu, K. Engelhorn, A. A. Correa, Y. Ping, J. W. Lee, L. J. Bae, D. Prendergast, R. W. Falcone, P. A. Heimann, J. Daligault, S. Gupta, S. H. Glenzer, J. W. Chan, T. Huser, S. Risbud, R. W. Lee, A. Ng, T. Ao, F. Perrot, M. W. C. Dharma-wardana, M. E. Foord, T. Ao, Y. Ping, G. M. Dyer, R. Ernstorfer, A. Mančić, B. I. Cho, T. G. White, Z. Chen, M. W. C. Dharma-wardana, F. Perrot, J. Vorberger, D. O. Gericke, T. Bornath, M. Schlanges, U. Reimann, C. Toepffer, A. Ng, P. Celliers, G. Xu, A. Forsman, D. Riley, Z. Lin, L. V. Zhigilei, V. Celli, B. I. Cho, S. Johnson, J. Hohlfeld, S. Wellershoff, J. Güdde, U. Conrad, H. Elsayed-Ali, T. Norris, M. Pessot, W. L. McMillan, G. Grimvall, E. Wohlfarth, L. B. Fletcher, M. G. Gorman, J. Gaudin, F. Dorchies, G. Paolo, D. Prendergast, and G. Galli, Sci. Rep. **6**, 18843 (2016).
[53] K. P. Migdal, N. A. Inogamov, Y. V. Petrov, and V. V. Zhakhovsky, Http://Arxiv.Org/Abs/1702.00825 (2017).
[54] A. P. Caffrey, P. E. Hopkins, J. M. Klopf, and P. M. Norris, Microscale Thermophys. Eng. **9**, 365 (2005).
[55] E. Beaurepaire, J. C. Merle, A. Daunois, and J. Y. Bigot, Phys. Rev. Lett. **76**, 4250 (1996).
[56] S.-S. Wellershoff, J. Hohlfeld, J. Güdde, and E. Matthias, Appl. Phys. A Mater. Sci. Process. **69**, S99 (1999).
[57] H. J. Zeiger, J. Vidal, T. K. Cheng, E. P. Ippen, G. Dresselhaus, and M. S. Dresselhaus, Phys. Rev. B **45**, 768 (1992).
[58] S. Chatterjee and D. K. Chakraborti, J. Phys. F Met. Phys. **1**, 638 (1971).
[59] P. B. Allen, T. P. Beaulac, F. S. Khan, W. H. Butler, F. J. Pinski, and J. C. Swihart, Phys. Rev. B **34**, 4331 (1986).
[60] R. H. M. Groeneveld, R. Sprik, and A. Lagendijk, Phys. Rev. Lett. **64**, 784 (1990).
[61] S. Edward, A. Antoncecchi, H. Zhang, H. Sielcken, S. Witte, and P. C. M. Planken, Opt.

Express **26**, 327654 (2018).
[62] J. Hohlfeld, S. S. Wellershoff, J. Gudde, U. Conrad, V. Jahnke, E. Matthias, J. Güdde, U. Conrad, V. Jähnke, and E. Matthias, Chem. Phys. **251**, 237 (2000).
[63] J. Kimling and D. G. Cahill, Phys. Rev. B **95**, (2017).
[64] M. Lejman, V. Shalagatskyi, O. Kovalenko, T. Pezeril, V. V. Temnov, and P. Ruello, J. Opt. Soc. Am. B **31**, 282 (2014).
[65] U. Ritzmann, P. M. Oppeneer, and P. Maldonado, Http://Arxiv.Org/Abs/1911.12414 (2019).
[66] M. Bonn, D. N. Denzler, S. Funk, M. Wolf, S.-S. Wellershoff, and J. Hohlfeld, Phys. Rev. B **61**, 1101 (2000).
[67] Y. Petrov, K. Migdal, N. Inogamov, V. Khokhlov, D. Ilnitsky, I. Milov, N. Medvedev, V. Lipp, and V. Zhakhovsky, Data Br. **28**, 104980 (2020).
[68] I. Milov, V. Lipp, N. Medvedev, I. A. Makhotkin, E. Louis, and F. Bijkerk, JOSA B **35**, B43 (2018).
[69] I. Milov, I. A. Makhotkin, R. Sobierajski, N. Medvedev, V. Lipp, J. Chalupský, J. M. Sturm, K. Tiedtke, G. de Vries, M. Störmer, F. Siewert, R. van de Kruijs, E. Louis, I. Jacyna, M. Jurek, L. Juha, V. Hájková, V. Vozda, T. Burian, K. Saksl, B. Faatz, B. Keitel, E. Plönjes, S. Schreiber, S. Toleikis, R. Loch, M. Hermann, S. Strobel, H.-K. Nienhuys, G. Gwalt, T. Mey, H. Enkisch, and F. Bijkerk, Opt. Express **26**, 19665 (2018).
[70] P. B. Allen, Phys. Rev. Lett. **59**, 1460 (1987).
[71] T. Ogitsu, A. Fernandez-Pañella, S. Hamel, A. A. Correa, D. Prendergast, C. D. Pemmaraju, and Y. Ping, Phys. Rev. B **97**, 214203 (2018).
[72] N. J. Hartley, P. Belancourt, D. A. Chapman, T. Döppner, R. P. Drake, D. O. Gericke, S. H. Glenzer, D. Khaghani, S. LePape, T. Ma, P. Neumayer, A. Pak, L. Peters, S. Richardson, J. Vorberger, T. G. White, and G. Gregori, High Energy Density Phys. **14**, 1 (2015).
[73] S. M. Oommen and S. Pisana, Https://Arxiv.Org/Abs/1910.05893 (2019).
[74] J. G. Fujimoto, J. M. Liu, E. P. Ippen, and N. Bloembergen, Phys. Rev. Lett. **53**, 1837 (1984).
[75] S. L. Daraszewicz, Y. Giret, H. Tanimura, D. M. Duffy, A. L. Shluger, and K. Tanimura, Appl. Phys. Lett. **105**, (2014).
[76] L. Waldecker, Electron-Lattice Interactions and Ultrafast Structural Dynamics of Solids (PhD Thesis), Freie Universität, Berlin, 2016.
[77] I. Lazanu and S. Lazanu, Astropart. Phys. **75**, 44 (2016).
[78] H. van Driel, Phys. Rev. B **35**, 8166 (1987).
[79] J. Li, A. Ciani, J. Gayles, D. A. Papaconstantopoulos, N. Kioussis, C. Grein, and F. Aqariden, Philos. Mag. **93**, 3216 (2013).
[80] T. G. White, J. Vorberger, C. R. D. Brown, B. J. B. Crowley, P. Davis, S. H. Glenzer, J. W. O. Harris, D. C. Hochhaus, S. Le Pape, T. Ma, C. D. Murphy, P. Neumayer, L. K. Pattison, S. Richardson, D. O. Gericke, and G. Gregori, Sci. Rep. **2**, 889 (2012).
[81] N. Medvedev, A. E. Volkov, and B. Ziaja, Nucl. Instruments Methods Phys. Res. Sect. B Beam Interact. with Mater. Atoms **365**, 437 (2015).
[82] B. Rethfeld, A. Kaiser, M. Vicanek, and G. Simon, Phys. Rev. B **65**, 214303 (2002).
[83] J. C. Tully, J. Chem. Phys. **93**, 1061 (1990).
[84] S. Hammes-Schiffer and J. C. Tully, J. Chem. Phys. **101**, 4657 (1994).